\RequirePackage{fix-cm}
\documentclass[smallcondensed]{svjour3}     
\smartqed  
\usepackage{graphicx,hyperref}
\usepackage[parfill]{parskip}    
\usepackage{amssymb}
\usepackage{amsmath}
\usepackage{multirow}
\journalname{Journal of Geodesy}
\begin{document}

\title{
Recovery of Bennu's Orientation for the OSIRIS-REx Mission
}

%

\subtitle{
Implications for the Spin State Accuracy and Geolocation Errors
}


\author{Erwan Mazarico         \and
        David D. Rowlands \and
        Terence J. Sabaka \and
        Kenneth M. Getzandanner \and
        David P. Rubincam \and
        Joseph B. Nicholas \and
        Michael C. Moreau
}


\institute{Erwan Mazarico \at
	      NASA Goddard Space Flight Center \\
              8800 Greenbelt Road , B34 W282 \\
              Greenbelt, MD 20771, USA
              Tel.: +1-301-614-6504\\
              \email{erwan.m.mazarico@nasa.gov}       
           \and
           David D. Rowlands \at
           NASA Goddard Space Flight Center, Greenbelt, Maryland, USA
           \and
           Terence J. Sabaka \at
           NASA Goddard Space Flight Center, Greenbelt, Maryland, USA 
           \and 
           Kenneth M. Getzandanner \at 
           NASA Goddard Space Flight Center, Greenbelt, Maryland, USA 
           \and 
           David P. Rubincam \at 
           NASA Goddard Space Flight Center, Greenbelt, Maryland, USA 
           \and 
           Joseph B. Nicholas \at
           Emergent Space Technologies Inc., Greenbelt,  Maryland, USA 
           \and 
           Michael C. Moreau \at 
           NASA Goddard Space Flight Center, Greenbelt, Maryland, USA
}

\date{Received: date / Accepted: date}

\maketitle

\begin{abstract}

The goal of the OSIRIS-REx mission is to return a sample of asteroid material from Near-Earth Asteroid (101955) Bennu.
The role of the navigation and flight dynamics team is critical for the spacecraft to execute a precisely planned sampling maneuver over a specifically-selected landing site. In particular, the orientation of Bennu needs to be recovered with good accuracy during orbital operations to contribute as small an error as possible to the landing error budget.
Although Bennu is well characterized from Earth-based radar observations, its orientation dynamics are not sufficiently known to exclude the presence of a small wobble.
To better understand this contingency and evaluate how well the orientation can be recovered in the presence of a large 1$^{\circ}$ wobble, we conduct a comprehensive simulation with the NASA GSFC GEODYN orbit determination and geodetic parameter estimation software. 
We describe the dynamic orientation modeling implemented in GEODYN in support of OSIRIS-REx operations, and show how both altimetry and imagery data can be used as either undifferenced (landmark, direct altimetry) or differenced (image crossover, altimetry crossover) measurements. We find that these two different types of data contribute differently to the recovery of instrument pointing or planetary orientation. When upweighted, the absolute measurements help reduce the geolocation errors, despite poorer astrometric (inertial) performance.
We find that with no wobble present, all the geolocation requirements are met. While the presence of a large wobble is detrimental, the recovery is still reliable thanks to the combined use of altimetry and imagery data.

\keywords{Orbit Determination \and Orientation \and Dynamics \and Asteroid}
\end{abstract}

\section{Motivation}
\label{sec:motivation}

In 2011, NASA selected a mission under the New Frontiers program to visit a near-Earth asteroid and return a pristine regolith sample to Earth (Lauretta et al., 2016). The Origins, Spectral Interpretation, Resource Identification, Security, Regolith Explorer, or OSIRIS-REx, is headed to (101955) Bennu under the direction of Dante Lauretta. The project is managed by NASA Goddard Space Flight Center (GSFC) in Greenbelt, Maryland and the spacecraft was built by Lockheed Martin in Denver, Colorado. OSIRIS-REx carries a rich payload of scientific instruments, supplemented by a Touch-and-Go (TAG) arm to collect up to 2 kg of regolith material.

	The primary challenge of the mission is to conduct a safe and successful sampling of the surface. In addition to enabling a good scientific understanding of the small asteroid ($\sim$ 250 m in radius) as a whole to select a landing site, the payload was designed to provide the data required to ensure TAG success. The collection of very-high-resolution datasets makes possible the selection of the best sampling site, both from a science interest standpoint and under strict engineering safety rules. Underpinning the safe execution of the sampling maneuvers, the dynamic environment of Bennu must be characterized precisely and robustly, and the spacecraft must then be navigated accurately to the TAG site. The OSIRIS-REx Flight Dynamics Team (FDT) is responsible for the navigation of the spacecraft. In combination with the Radio Science Working Group and the Altimetry Working Group, the dynamic environment of Bennu, such as the gravity field, shape, ephemeris, and orientation of the asteroid, will be estimated in the mission phases leading up to sampling.

	Although no non-principal-axes rotation was identified from repeated ground-based radar imaging \cite{nolan2013}, the presence of a wobble could compromise, or at least complicate, the operation planning. In pre-launch preparations, several simulations were jointly executed by the various groups and institutions forming the FDT. Here, we present the results of additional simulations performed at NASA GSFC in order to address the recovery of a wobble on Bennu, and its implications for the TAG maneuver. We used the GEODYN II orbit determination and geodetic parameter estimation software developed and maintained at NASA GSFC \cite{pavlis2013}. We focused on the orbital configuration expected during the so-called Orbital Phase B, which is important because it will be the staging orbit for the TAG maneuver.

	The outline of this manuscript is as follows. In Section~\ref{sec:background}, we describe the asteroid Bennu and the OSIRIS-REx mission. In Section~\ref{sec:orientation}, we present important concepts related to planetary orientation definition and representation, introducing the analytical and dynamical approaches to model it for the purposes of Orbit Determination (OD). We also discuss the case of Bennu, including the characteristics of its potential wobble. We then consider practical aspects related to OD, namely the linearity and stability of orientation estimation, the pertinence of the dynamical approach, and the limits to the simplicity of actual asteroid rotation. In Section~\ref{sec:od}, we explain in detail our Orbit Determination methodology. We introduce the models, measurement types, and simulation capabilities implemented in our GEODYN that are relevant to this study. We also discuss the strengths of the image-based and altimetric measurement types used in our simulation. In Section~\ref{sec:wobblerecovery}, we focus on the simulation of wobble recovery, building the case with small side simulations for the approach selected in our comprehensive simulation. We present the results of this realistic, full-up simulation, particularly regarding the geolocation performance when different data weights are used. We then discuss the reasons why it is difficult to recover asteroid orientation with sub-meter geolocation performance. Finally, we summarize and conclude in Section~\ref{sec:summary}. An appendix (Section~\ref{sec:appendix}) provides details on the computation and estimation of the planetary orientation parameters through a dynamical approach.

\section{Background}
\label{sec:background}

\subsection{Bennu}
\label{sec:bennu}

The asteroid (101955) Bennu, originally known as 1999RQ36, is a potentially hazardous asteroid whose orbit crosses Earth's. Considering it has not been visited by spacecraft yet, numerous physical properties of Bennu are known with good accuracy. These are based on surveys performed by the large radio telescope at Arecibo \cite{nolan2013}. For the purposes of science preparation and mission planning, the OSIRIS-REx project established reference parameters informed by existing scientific observations \cite{nolan2013}.

Bennu is a B-type asteroid with an average radius of $\sim$ 250 m and an equatorial radius of $\sim$ 275 m. Ground-based radar observations were only weakly sensitive to the polar radius, which is largely unconstrained, save for considerations of rotational stability given the moments of inertia predicted with a uniform density. The rotation period is well-known, $\sim$ 4.3 hours. From radar and lightcurve analysis, no wobble rotation was detected, but the uncertainty in spin axis position is sufficiently large ($\pm$ 4$^\circ$) that it cannot be excluded presently. The ephemeris of Bennu is known to a few kilometers \cite{chesley2014} and will be refined from spacecraft ranging data during the OSIRIS-REx mission.
The mass of Bennu was obtained from a novel combination of radio astronomy and infrared observations, by observation of the Yarkovsky effect and of the surface dielectric properties. The gravitational parameter (GM, with G the gravitational constant and M the mass) is estimated to be 5.2 $\pm$ 0.9 m$^3$s$^{-2}$. The low bulk density ($\rho \approx$ 1260 kg/m$^3$) indicates that Bennu is likely a rubble-pile with a macro-porosity of $\sim$ 40\%.

\begin{figure}
\centering\includegraphics[width=4.5in]{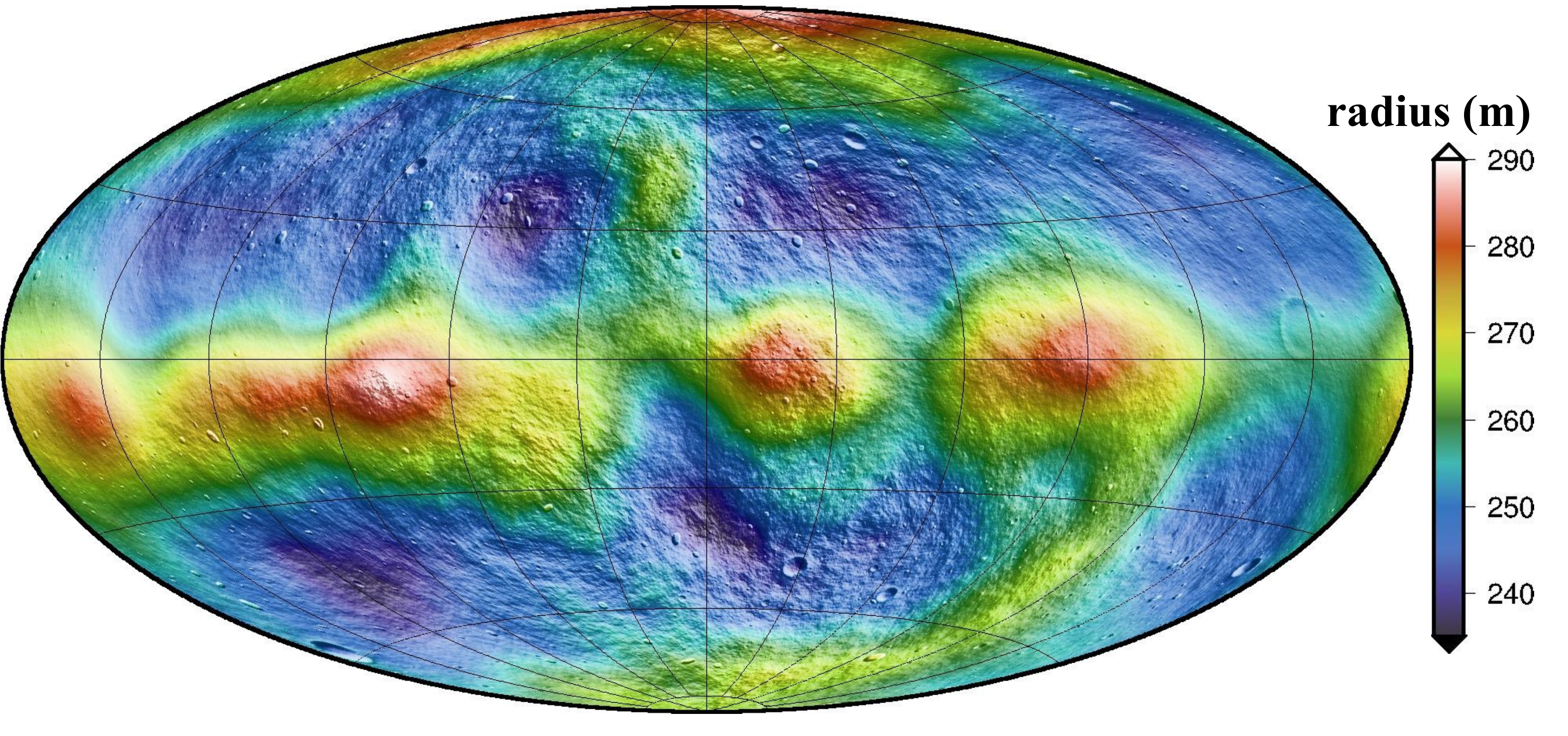}
\caption{High-resolution artificial shape of Bennu used in the simulation. Created by R. Gaskell, it is in agreement with the shape derived from ground-based radar data. Mollweide projection centered on (0$^{\circ}$E,0$^{\circ}$N).}
\label{fig1}
\end{figure}

\begin{figure}
\centering\includegraphics[width=2.5in]{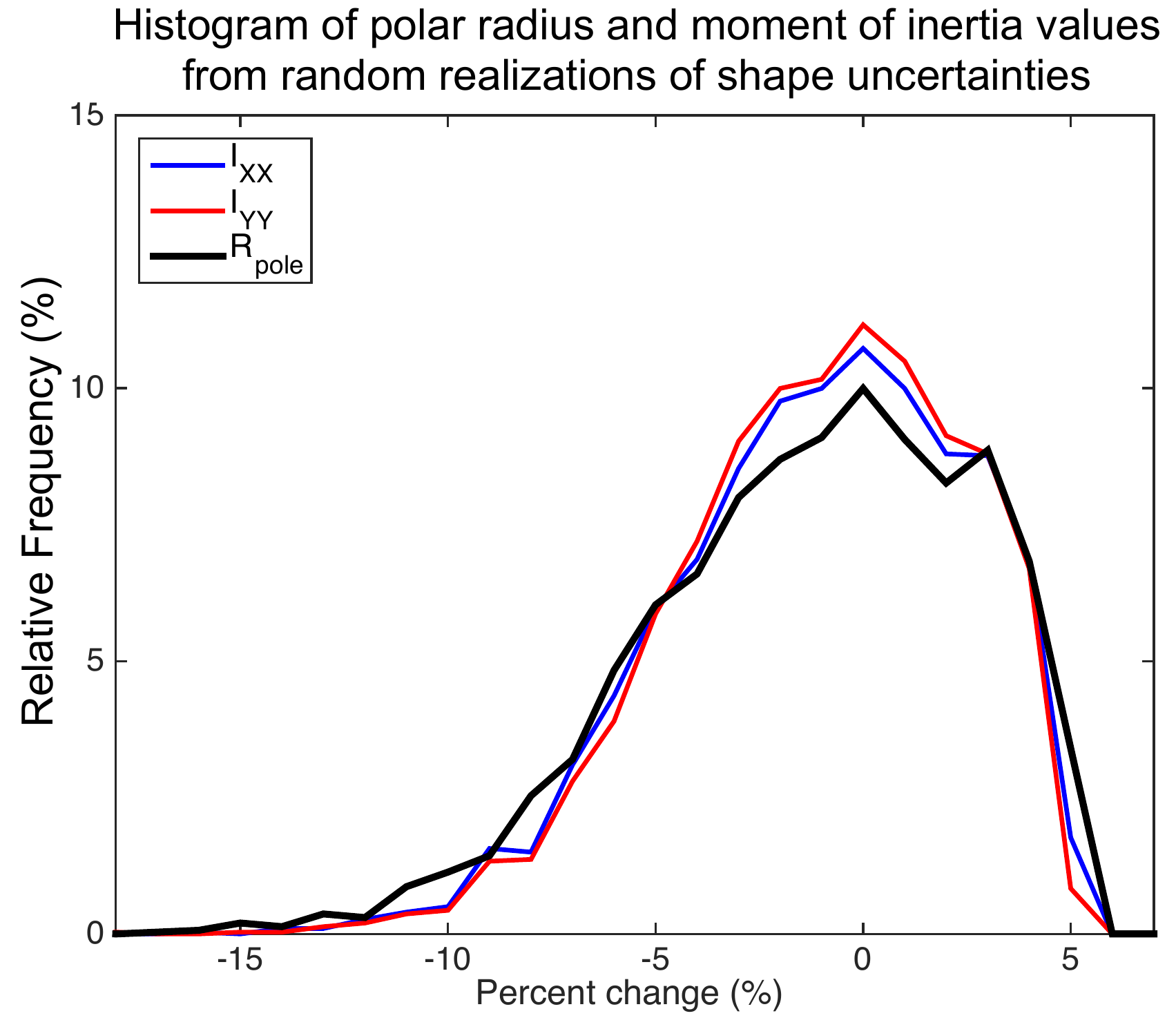}
\caption{Distribution of the relative changes in polar radius and derived moments of inertia $I_{xx}$ and $I_{yy}$, from the best-fit parameters. The polar radius is consistent with uncertainties in the radar-derived shape obtained by \cite{nolan2013}. The sharp drop-off at larger values is due to dynamical stability considerations.}
\label{fig2}
\end{figure}

Moments of inertia and gravity field models can be prescribed assuming a uniform density and a shape, such as best-fit ellipsoid to the radar data. A reference asteroid shape model consistent with Bennu's measured properties was developed by R. Gaskell (Planetary Science Institute) for use during mission planning and simulations (Figure~\ref{fig1}). In particular, this shape is used to simulate the image and altimetry data acquired by OSIRIS-REx. We derived the inertia tensor for that complex and realistic artificial body shape (Eq.~\ref{eq1}), which will be used in the simulations presented below, under the assumption of uniform density. A more complex internal structure is of course possible, potentially as a result of the spin history as studied by \cite{hirabayashi2015}, but is not directly relevant to our work, given we estimate the inertia tensor and do not rely on any internal density assumption to obtain our solution.  To assess the possible range of those parameters, we used the shape uncertainties from \cite{nolan2013} and recomputed the inertia tensor. As explained by \cite{nolan2013}, because of the viewing geometry during the radar observations, the radar data are relatively insensitive to 5\% errors in polar radius, with values of 10\% allowable within uncertainty. The equatorial radii are better constrained, to about 2\% ($\pm 10$ m). As a result, the recovered shape's flattening is loosely constrained to 4±8\%. We created a sample of 3,000 Gaussian-distributed errors on the polar radius with a standard deviation of 5\%, after discarding values that would yield a polar radius larger than the equatorial radius, which \cite{nolan2013} consider unlikely for dynamical stability reasons. Figure~\ref{fig2} shows the resulting distributions of the variations in polar radius, and $I_{xx}$ and $I_{yy}$ moments of inertia. ($I_{zz}$ does not change with our simple scaling of the Z coordinates of the shape model vertices.) \cite{nolan2013} also discuss a possible bias in the overall size of Bennu of up to 2\%. Because the rotational dynamics in the torque-free case only depends on the ratios between the moments of inertia (Section~\ref{sec:smallperiodic} and Appendix in Section~\ref{sec:appendix}), the orientation of Bennu is insensitive to change in scale (and in bulk density), and we do not consider it further.

\subsection{OSIRIS-REx}
\label{sec:osiris}

The OSIRIS-REx spacecraft carries a capable suite of high-resolution instruments \cite{lauretta2016} designed to characterize Bennu in unprecedented detail, down to decimeter-scale globally. Of interest for our study and the geodetic aspect of the mission, the OSIRIS-REx Laser Altimeter (OLA; \cite{daly2016}) contributed by the Canadian Space Agency (CSA) will provide ranging information to the surface between 0.04 and 9 km altitude. A scanning mirror and high firing rate allow a large number of altimetric returns to be collected in short periods and to form three-dimensional raster images. Two navigation cameras (NavCams) will acquire high-resolution images of the asteroid in order to support optical navigation (OpNav), necessary to ensure meeting the position knowledge and prediction requirements during operations. The lack of an optical bench on OSIRIS-REx means that the pointing of its high-resolution instruments needs to be calibrated regularly, through estimation during OD for instance.

After nearly two years of cruise, the OSIRIS-REx spacecraft will reach Bennu in August 2018. Because of the very weak gravity environment, solar radiation pressure substantially affects the long-term evolution and stability of spacecraft orbits around Bennu. A terminator is preferred for mission operations to avoid unstable trajectories that would be difficult to recover in the case of a prolonged anomaly (e.g., Scheeres, 2012).

	During final approach to Bennu, an initial shape model will be developed. The subsequent Preliminary Survey Phase will help determine Bennu's mass at the 1\% level from DSN radio tracking during 3 hyperbolic passes. It will also provide a 75-cm resolution shape model with an accuracy requirement of 1 m. During the next phase, Orbital A, OSIRIS-REx will be placed in a 1.5-km radius circular terminator orbit in order to obtain the imagery necessary to construct a global high-resolution shape model through stereo-photoclinometry \cite{gaskell2008}. NavCam images acquired at a cadence of 4 hours will start being used for surface landmark OpNav, as opposed to stellar OpNav. A detailed survey phase will characterize potential landing sites, and improve the shape model to 35-cm resolution and accuracy better than 75 cm. Orbital Phase B will precede reconnaissance passes over four sites at a range of 225 meters and two TAG passes (one for rehearsal, and one for actual sample collection). Orbital Phase B is a month-long 1-km radius circular terminator orbit with one dedicated 9-day period of near-continuous DSN tracking for radio science and gravity field determination. OLA data are also collected to construct a shape model independent of those derived from images.

	The Orbital Phase B orbit is important for the mission given it is the staging orbit for critical maneuvers (reconnaissance, TAGs). As such, insight on the quality of the spacecraft trajectory prediction capability and of the Bennu orientation estimation are required to ensure successful TAG within 25 meters of a chosen sampling site \cite{berry2013}. One particularly relevant requirement for the Flight Dynamics team is that the error of TAG location contributed by the mismodeling of Bennu orientation should be less than 1 m after 28 hours. 
	
\section{Orientation and Wobble}
\label{sec:orientationwobble}

\subsection{Planetary Orientation}
\label{sec:orientation}

The transformation between the Bennu-fixed frame and inertial space (defined by background stars, such as the International Celestial Reference Frame, or ICRF; \cite{ma1998}) is required for navigation and science. For instance, such rotation matrices are needed to geolocate the data collected by various instruments at different times onto Bennu's surface. With the very high resolution of the datasets to be acquired by OSIRIS-REx, a good precision ($<$1 m) is desirable, or even required.

\begin{figure}
\centering\includegraphics[width=4.5in]{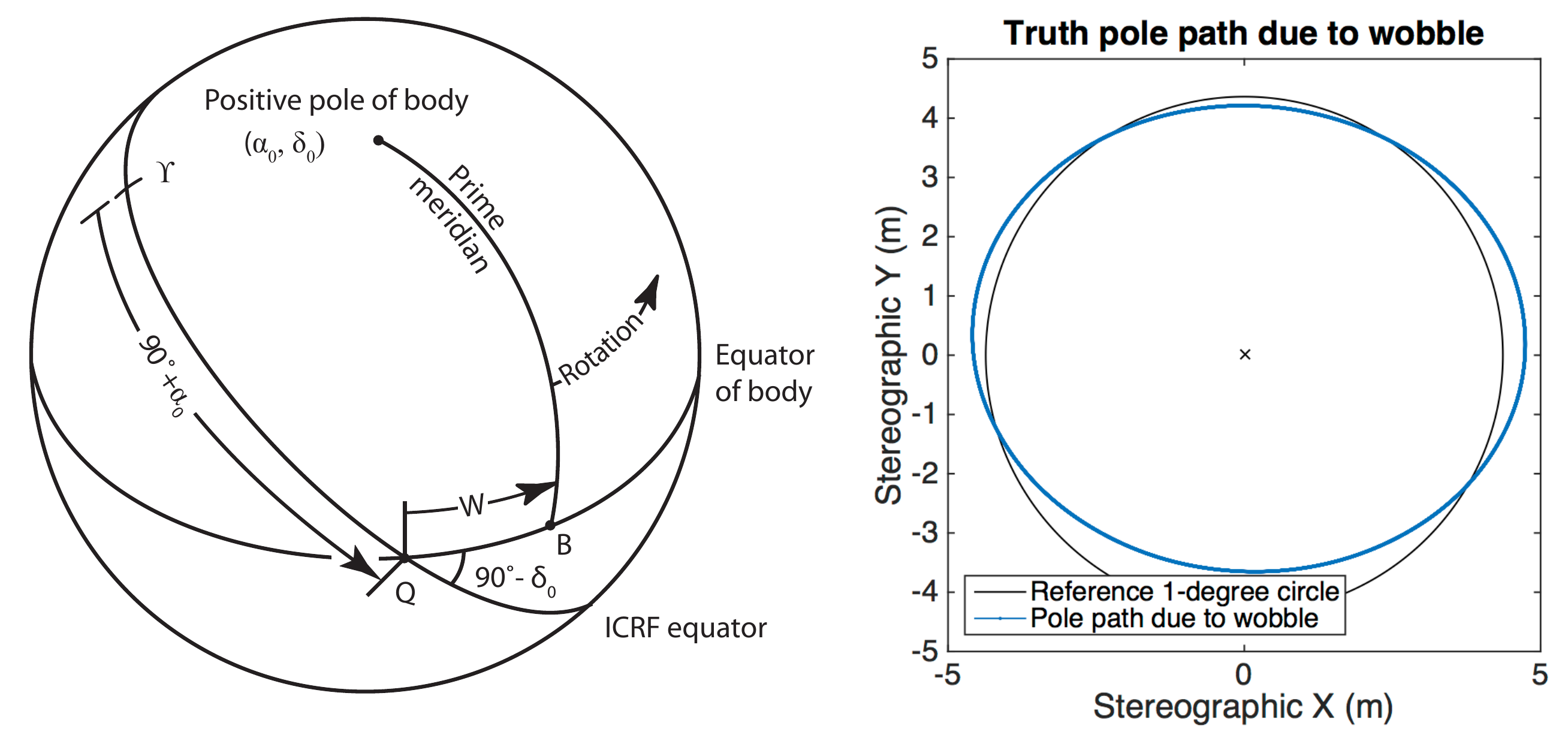}
\caption{(a, left) Schematic description of the IAU definition of the three angles defining body orientation (Fig. 2 from \cite{archinal2010} with permission from author). (b, right) Path of the true pole of Bennu due to wobble, computed for the nominal inertia tensor and shown in polar stereographic projection.}
\label{fig3}
\end{figure}

	Following IAU convention, a rotation transformation is given by three angles (Figure~\ref{fig3}a): right ascension (RA; $\alpha$), declination (DEC; $\delta$), and prime meridian location (W). At any epoch, the transformation between inertial frame and Bennu orientation is defined by three such angles. All three can have time variations. The time history of W is the spin, mainly secular. The time history of $\alpha$ and $\delta$ are the precession and wobble. We do not consider free librations in this work, as they are unlikely to survive for an extended duration on Bennu.

	The J2000 coordinate system is the reference inertial coordinate system in GEODYN. By definition, the Z axis is the pole of the Earth at the J2000 epoch (Jan 1, 2000 at noon). The X-Y plane is perpendicular to the Z axis (e.g., Earth Equator at J2000 epoch), and the X axis is along the intersection of the X-Y plane and the plane of the Earth orbit.

To give the orientation of the Bennu pole (Z axis of Bennu) and of the Bennu Equator with respect to J2000, two angles are required: $\alpha$ (longitude of Bennu Pole at J2000) and $\delta$ (latitude of Bennu Pole at J2000). W is the angle between the Bennu prime meridian and the IAU vector Q, the intersection of Bennu's equator and the J2000 X-Y plane. With this third angle, the location of any point on Bennu's surface can be related to the inertial J2000 coordinate system.

	Though not arbitrary, the time history of those three angles can vary in complexity. The simplest rotation case corresponds to fixed $\alpha$ and $\delta$ values and W linear in time (constant derivative $\dot{\text{W}}$, or spin rate). Major planets are in this state within good approximation. Longitudinal librations such as that of Mercury \cite{margot2009} can be represented by an additional series of periodic terms on W. Simple nutation and precession can also be represented by analytical periodic terms on $\alpha$ and $\delta$.

	A wobble is characterized by the three $\alpha$, $\delta$, and W angles all having non-zero periodic terms. The Earth has a wobble (polar motion), first recognized by Chandler, and considerable effort is spent monitoring it to provide accurate Earth Orientation Parameters \cite{seidelmann1982} used in many geodetic applications.  A wobble state means that there is a rotation about all three body-fixed axes, that the position of the instantaneous spin axis is changing in the Bennu-fixed frame, and that this instantaneous spin axis would trace out an ellipse centered at Bennu's body-fixed Z axis with a given period. A one-degree wobble means that the semi-major axis of the spin axis ellipse is about 1 degree. Although the wobble is driven by the instantaneous spin axis, one really needs the time series of the three $\alpha$, $\delta$, and W angles to relate inertial J2000 and Bennu body-fixed frames.

\subsection{How can orientation be represented?}
\label{sec:orientationrepresentation}

The simplest way to represent orientation is analytical, whereby a series of constant, linear, and period terms of given frequencies are prescribed or determined for each of the $\alpha$, $\delta$, and W angles. Each periodic effect is modeled as the sum of a cosine term and of a sine term, whose amplitudes yield a total amplitude and a phase. A wobble can be approximated by using such terms of that given wobble period for both $\alpha$ and $\delta$.

It is also possible to determine these three angles at every time step of a numerical integration of the differential equations describing the `force model' of the body dynamics \cite{goldstein1965}. This dynamical approach requires twelve parameters to derive a time series of orientation: six for the initial orientation state ($\alpha_0$, $\delta_0$, W$_0$, $\dot{\alpha}_0$, $\dot{\delta}_0$, $\dot{\text{W}}_0$) and six for the inertia tensor $\mathbf{I_{c}}$ ($I_{xx}$, $I_{yy}$, $I_{zz}$, $I_{xy}$, $I_{xz}$, $I_{yz}$). This approach is rather complex to implement in orbit determination software, but affords more capability and flexibility in case a wobble is actually discovered at Bennu. An Appendix (Section~\ref{sec:appendix}) shows in detail the derivation of the variational equations of the orientation with respect to the twelve parameters of the initial orientation state. As part of the OSIRIS-REx mission preparations, we implemented this dynamical approach into our GEODYN software.
In other words, at a given epoch we interpolate the $\alpha$, $\delta$ and W parameters as part of a numerical integration that starts at the same initial epoch as the orbit state. We also interpolate partial derivatives of the $\alpha$, $\delta$ and W angles with respect to the 6 initial state and the 6 inertia tensor parameters from a numerical integration. These partial derivatives are used to get the partial derivatives of the measurements with respect to the 6 initial state and the 6 inertia tensor parameters. Some the particulars of the construction of measurement partial derivatives will be discussed in Section~\ref{sec:obstypespartials}.

\subsection{Wobble in the case of Bennu}
\label{sec:wobblebennu}

Given current knowledge of Bennu's bulk density, shape, and orientation (Section \ref{sec:bennu}), reasonable estimates for the moments of inertia can be computed and used as a priori for simulations (Eq. \ref{eq1}). We compute the inertia tensor in the body-fixed cartesian frame $\mathbf{I_{c}}$ from of polyhedral shape model following \cite{dobrovolskis1996}.

\begin{equation}
\mathbf{I_{c}}=
\left({
\begin{array}{ccc}
1.752 \times 10^{15} & 7.596 \times 10^{10} & -2.448 \times 10^{11} \\
7.596 \times 10^{10} & 1.820 \times 10^{15} & 3.457 \times 10^{11} \\
-2.448 \times 10^{11} & 3.457 \times 10^{11} & 1.968 \times 10^{15}
\end{array}}\right) \text{kg.m}^{2}\label{eq1}
\end{equation}

Note that the off-diagonal elements are several orders of magnitude smaller than the diagonal elements, indicating that with a uniform density, the radar-derived shape is nearly in a principal-axes frame.
The wobble period can be derived from the moments of inertia values, analytically with the following formula if $I_{xx} \sim I_{yy}$ and torque is negligible:
\begin{equation}
\omega_{\text{wobble}} = \frac{I_{xx}-I_{zz}}{I_{xx}} \omega_{\text{spin}}
\label{eq2}
\end{equation}

Using the average of $I_{xx}$ and $I_{yy}$ in Eq. \ref{eq2}, we find a wobble period of approximately 42.1 hours.

To obtain the `truth' state of the 12 parameters for our 1-degree wobble simulation, we started with the 9 parameters that are better known (Eq.~\ref{eq1}, Eq.~\ref{eq3}) together with setting W$_{0}$ to zero, and then we searched for the $\dot{\alpha}_{0}$ and $\dot{\delta}_{0}$ parameters that yield an instantaneous rotation axis with a latitude nearly constant around $\sim$ 89$^{\circ}$ on Bennu. The resulting values are in Equation~\ref{eq4}. We also found from the numerical integration that this state (Eq.~\ref{eq1}, Eq.~\ref{eq3}, Eq.~\ref{eq4} plus W$_{0}$=0)  produces a wobble with a period of 43.2 hours, close to the expected period from the analytical formula given in Eq.~\ref{eq2} considering the $\sim3.7\%$ difference between $I_{xx}$ and $I_{yy}$.

\begin{equation}
\begin{array}{rcl}
\alpha_{0} &=& 86.5^{\circ} \\
\delta_{0} &=& -65^{\circ} \\
\dot{\text{W}}_{0} &=& 2014^{\circ}/\text{day}
\end{array}
\label{eq3}
\end{equation}

\begin{equation}
\begin{array}{rcl}
\dot{\alpha}_{0} &=& 16.28^{\circ}/\text{day} \\
\dot{\delta}_{0} &=& -36.62^{\circ}/\text{day}
\end{array}
\label{eq4}
\end{equation}

Figure~\ref{fig3}b shows the path of the Bennu spin pole in a north polar stereographic projection. The deviation from a circular path is due to the non-equal $I_{xx}$ and $I_{yy}$ moments of inertia.

The period of the $\alpha$, $\delta$ and W periodic terms is dictated by the overall spin rate $\dot{\text{W}}$ (360 degrees in 4.3 hours) and the the rate of the instantaneous spin axis motion within Bennu (360 degrees every 43.2 hours). The angular rates are additive, so in this case the period of the orientation angles is $\sim$ 3.9 hours. In this 1-degree wobble example, the amplitudes of the $\alpha$, $\delta$, and W periodic terms are $\sim2.1^{\circ}$, $\sim0.9^{\circ}$, and $\sim1.9^{\circ}$.

\subsection{Practical Considerations for Orbit Determination}
\label{sec:practicalwobble}

One important consideration when performing orbit determination and reducing tracking data is the linearity and stability of parameter estimation. The Appendix (Section~\ref{sec:appendix}) details the parametrization of the dynamical model, and how it integrates in the orbit determination software.
The partial derivatives of the numerically integrated $\alpha$, $\delta$ and W angles with respect to the moment parameters are themselves numerically integrated. They are nonlinear and starting at about 3 days after the initial epoch, the partials  begin to fail to predict the change in the angles from a change in the moments. This practically limits the length of estimation arcs over which data can be analyzed. On the other hand, the analytical approach requires knowledge of the period to be used for the cosine and sine terms for the  angles. If the period is to be estimated it is also a nonlinear parameter.

\begin{figure}
\centering\includegraphics[width=4.5in]{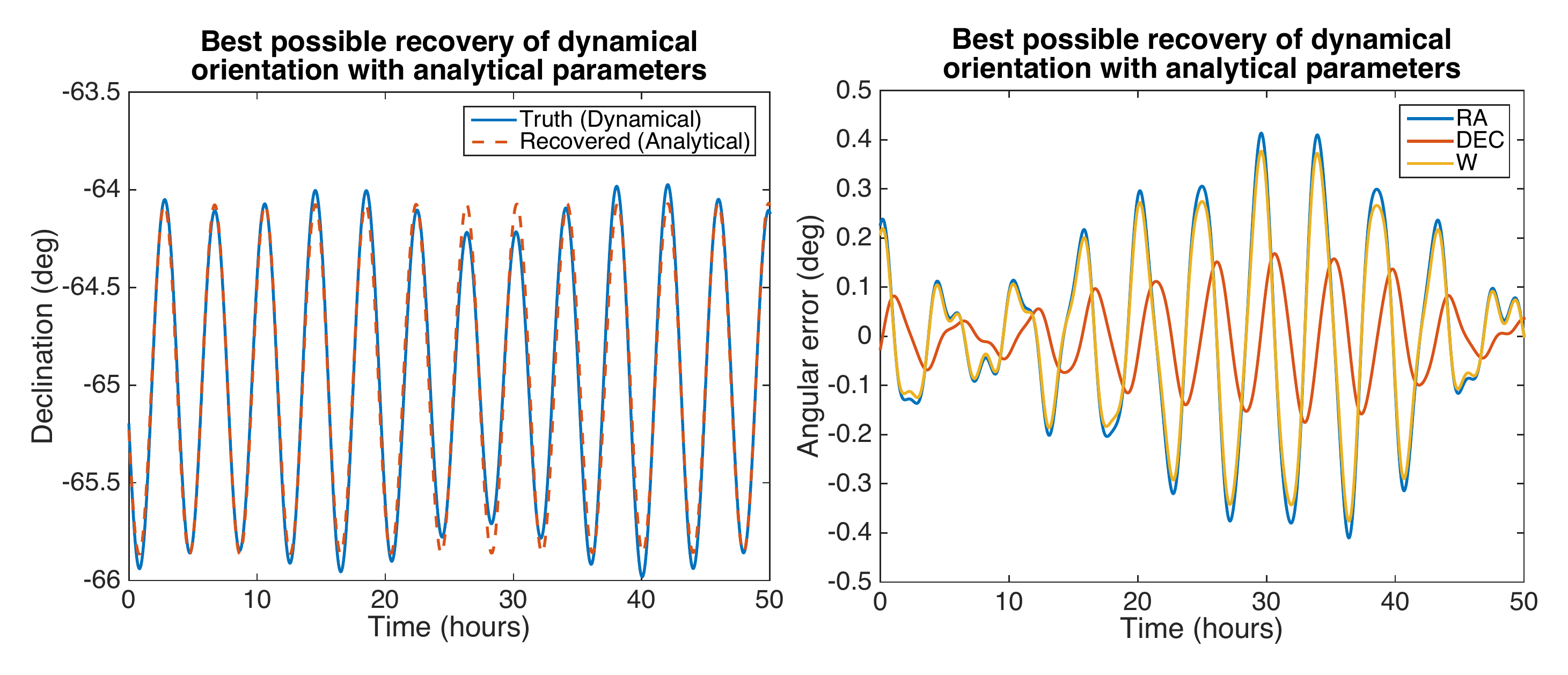}
\caption{(left) Time series of the true and analytically recovered declination orientation angle of Bennu. (right) Time series of the errors in the recovered values of the three orientation angles.}
\label{fig4}
\end{figure}

	Most critically, can the analytical approach correctly represent the wobble of a complex body with a non-diagonal inertia tensor? We address this question by performing a small, self-contained simulation over a 3-day time span. We first generate a truth time series using our dynamical model, and then estimate $\alpha$, $\delta$, and W analytical parameters using instantaneous angle measurements (Figure~\ref{fig4}). Those `measurements' are values recording the orientation angles themselves at a given time, and are thus of the realm of thought experiment given no instrument would ever be able to record those directly. But they represent `measurements' with perfect observability and thus yield the most optimistic inversion performance. We add no noise to those perfect, ideal measurements of orientation (which again would never be directly accessible in reality), and do not adjust the wobble frequency, fixed at its truth value. Despite these optimistic assumptions, we find that the analytical formulation can only approximate the wobble orientation, and yield a typical geolocation error on the order of one meter (Table~\ref{tab1}). A two-day prediction period after the three-day solution period does not show degraded results, so this illustrates a fundamental inability of the analytical model to capture the complex wobble dynamics.

\begin{table}
\caption{RMS and maximum errors in orientation angles ($\alpha$, $\delta$, W; in degrees) and in position (at 3 latitudes; in meters) due to the use of an analytical orientation model compared to a dynamical orientation model.}
\label{tab1}
\centering\begin{tabular}{|l|ll|ll|ll|ll|ll|ll|}
\hline
Errors & \multicolumn{2}{c|}{$\alpha$ ($^{\circ}$)} & \multicolumn{2}{c|}{$\delta$ ($^{\circ}$)} & \multicolumn{2}{c|}{\text{W} ($^{\circ}$)} & \multicolumn{2}{c|}{$\lambda = 0^{\circ}$ (m)} & \multicolumn{2}{c|}{$\lambda = 45^{\circ}$ (m)} & \multicolumn{2}{c|}{$\lambda = 90^{\circ}$ (m)} \\
  & RMS  & Max. & RMS  & Max. & RMS  & Max. & RMS  & Max. & RMS  & Max. & RMS  & Max.  \\
\hline
Solution (0-36 h) & 0.181 & 0.413 & 0.076 & 0.175 & 0.164 & 0.376 & 1.07 & 2.40 & 0.83 & 2.19 & 0.47 & 0.81 \\
\hline
Prediction (36-50 h) & 0.200 & 0.457 & 0.084 & 0.186 & 0.181 & 0.415 & 1.20 & 2.58 & 0.91 & 2.29 & 0.52 & 0.89 \\
\hline
\end{tabular}
\end{table}

In a simulation described further below (see Section~\ref{sec:pointing}), we will use the 1$^{\circ}$ wobble case, which is rather large; it is not expected at Bennu, but not excluded from ground-based data. 
In principle, the rotation can be perfectly quiet, in the principal axes frame (where the inertia tensor is diagonal). However, there are practical limits on how quiet it will appear given the imperfect choice of the Bennu-fixed reference frame will not be exactly aligned with the principal axes frame.
One interesting question is how small of a wobble one can expect to have to deal with at Bennu, given our imperfect knowledge of the inertia tensor even after arrival.
Since Bennu is rotating fast ($>$2000$^{\circ}$/day), there is a limit of how `quiet' (simple) its rotation will appear in a given Bennu-fixed frame, due to possible off-diagonal moments of inertia in that frame (e.g., Eq.~\ref{eq1}).
In a complementary simulation to that discussed above, we produced a 3-day time series of orientation angles based on a simple analytical orientation model (constant $\alpha$, $\delta$, and $\dot{\text{W}}$; with no periodic terms), and we used the dynamical model approach to adjust the initial orientation state to best match the orientation angle time series, while retaining the moments fixed to Bennu's nominal values (Table~\ref{tab1}). The adjustment was allowed to converge over three iterations, and resulted in the discrepancies listed in Table~\ref{tab2}. This shows that a geolocation of 0.3-0.6 m at a minimum is expected even on a quiet Bennu due to the off-diagonal moments of inertia, if analytical parameters are to be used. Indeed, there are no errors other than how well a set of simple analytical orientation parameters can match what the dynamical model dictates.  If better position requirement is needed, either additional analytical periodic terms have to be estimated, or the dynamical model should be used. The periodic parameters require a period be prescribed, but it is unknown as it depends on the moments of inertia. On the other hand, with the dynamical model, the moments are very non-linear and restrict the temporal extent of the estimation period.

\begin{table}
\caption{RMS and maximum errors in geolocated position (at 3 latitudes; in meters) due to the use of an analytical orientation model compared to a dynamical orientation model, in the case of a quiet Bennu.}
\label{tab2}
\centering\begin{tabular}{|l|ll|ll|ll|}
\hline
& \multicolumn{2}{c|}{$\lambda = 0^{\circ}$} & \multicolumn{2}{c|}{$\lambda = 45^{\circ}$} & \multicolumn{2}{c|}{$\lambda = 90^{\circ}$} \\
  & RMS & Max. & RMS & Max.  & RMS & Max. \\
\hline
 Geolocation Errors (m)  & 0.66 & 0.91 & 0.51 & 0.83 & 0.29 & 0.30 \\
\hline
\end{tabular}
\end{table}

\section{Orbit Determination}
\label{sec:od}

\subsection{GEODYN}
\label{sec:geodyn}

\begin{figure}
\centering\includegraphics[width=4.5in]{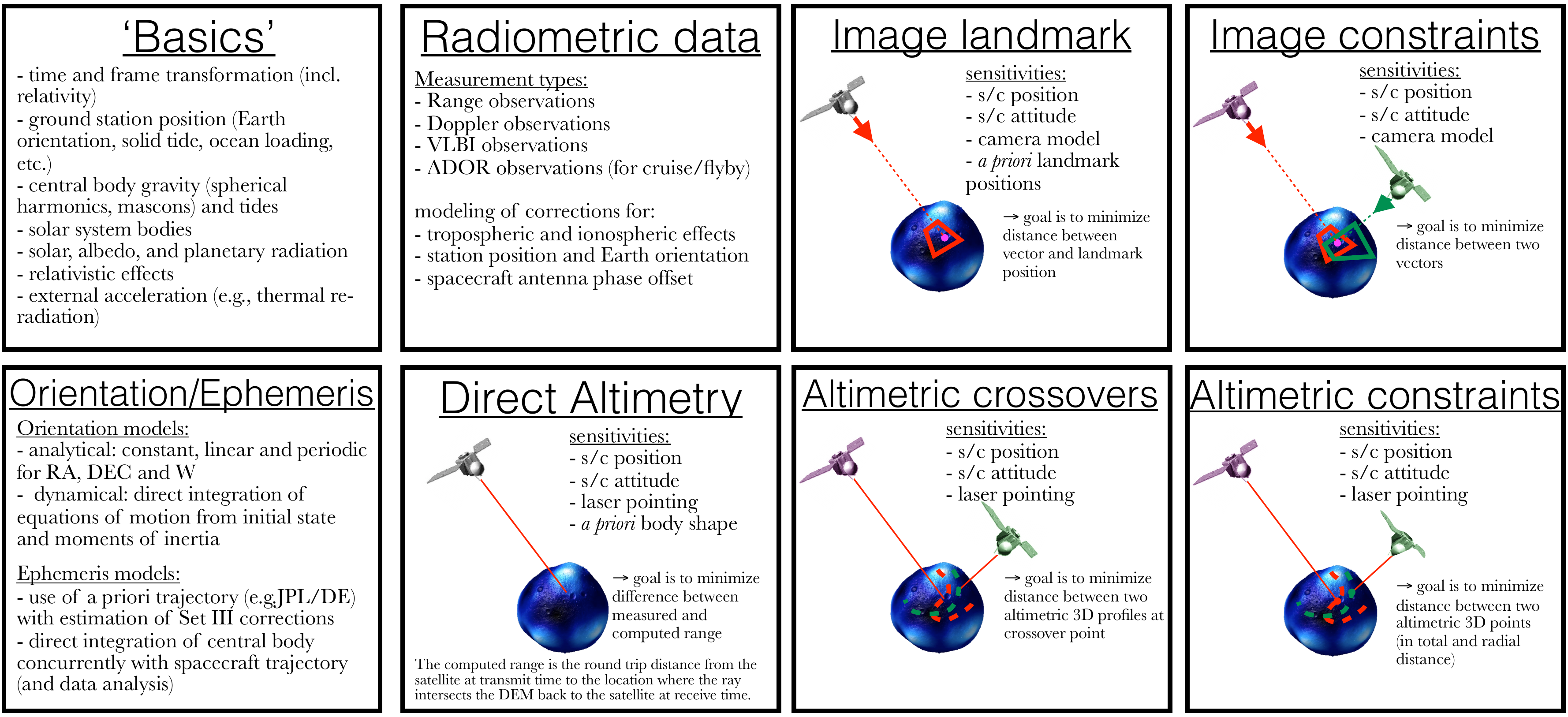}
\caption{Summary of the main models implemented in the GEODYN II software in order to integrate spacecraft trajectory, model tracking observations, and estimate geophysical parameters.}
\label{fig5}
\end{figure}

In support of the OSIRIS-REx mission, we will analyze radio tracking as well as image and altimetry data with the NASA GSFC GEODYN II program \cite{pavlis2013}. GEODYN is an orbit determination and geodetic parameter estimation software, which integrates spacecraft trajectory using a set of force models, models the observations using a set of measurement models, and performs batch least-squares to minimize the observation residuals while estimating model parameters. Since its inception as a tool focused on Earth-orbiting satellites, GEODYN's capabilities have grown, in particular for planetary science applications. It benefits from state-of-the-art models developed for high-accuracy Earth-focused geodetic missions, and it can now process a large number of measurement types. Figure~\ref{fig5} summarizes the main force and measurement models implemented in GEODYN that will be used during the OSIRIS-REx mission.

\subsubsection{General Models}
\label{sec:generalmodels}

GEODYN processes batches of data over independent time periods, called `arcs'. The arc duration is selected by the user, depending on data coverage, spacecraft maneuvers, and the specific goals of the OD performed. Arcs can be as short as a few minutes for local gravity studies \cite{luthcke2008,goossens2014} and as long as several weeks if small, sensitive parameters need to be adjusted  \cite{zuber2000,mao2016}. The most typical duration in planetary gravity studies is 1 to 5 days \cite{lemoine2001,mazarico2012,mazarico2014}, but analysis of NEAR data was performed over mission phases that could last several weeks \cite{zuber2000,konopliv2002}.

\begin{figure}
\centering\includegraphics[width=3in]{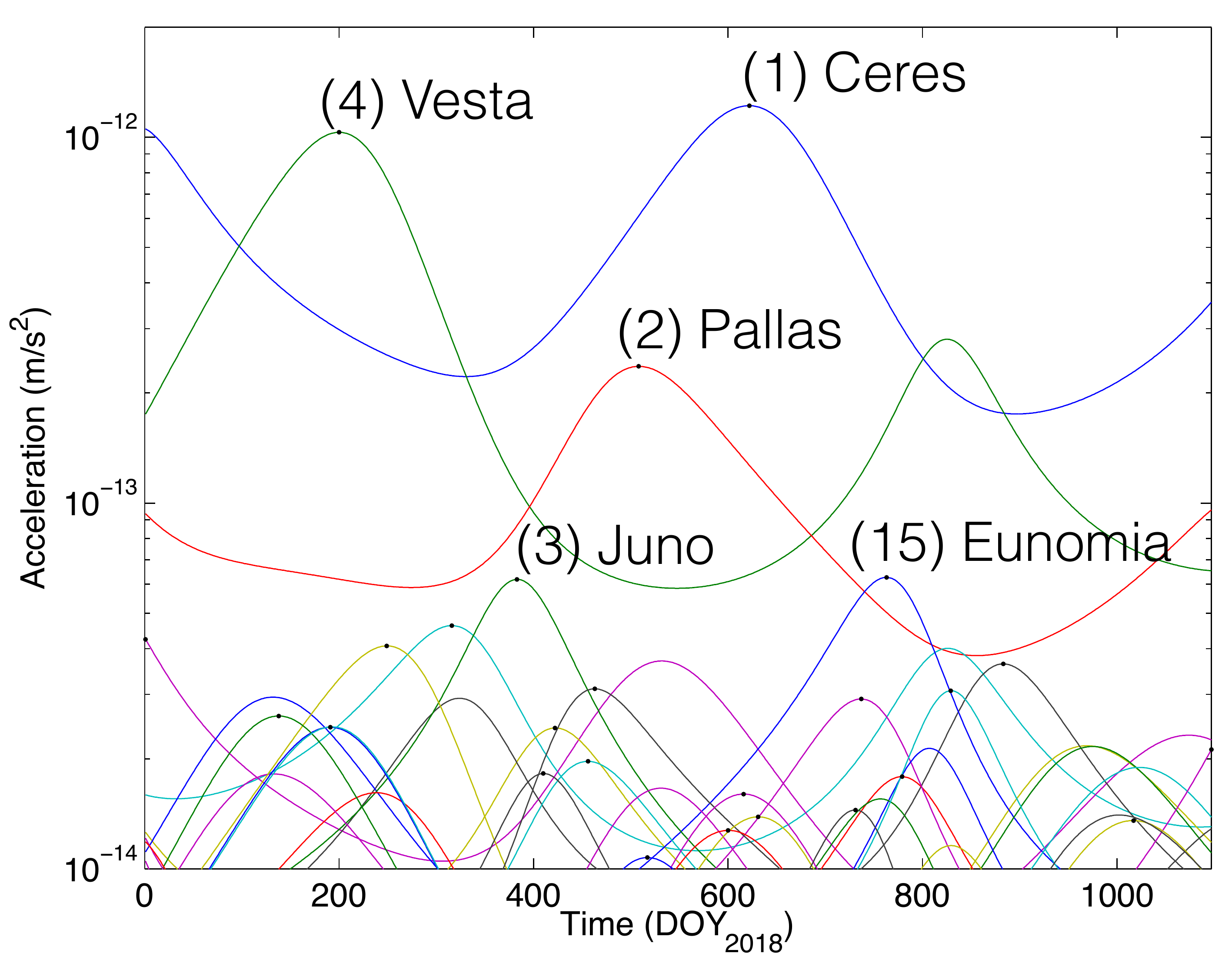}
\caption{Accelerations induced by other asteroids onto Bennu during the OSIRIS-REx mission span. These perturbations will be accounted for during the integration of Bennu's ephemeris.}
\label{fig6}
\end{figure}

	For its integration of spacecraft trajectory over an arc, starting from an initial state (position and velocity), GEODYN accounts for a variety of forces. Gravitational accelerations are computed using a full spherical harmonics representation of the gravity field of the orbited body, and using point mass approximation for third-body perturbations from major planetary bodies and a selectable number of asteroids. We use the Solar System ephemerides developed at JPL (DE432; \cite{folkner2014}). GEODYN can directly use these ephemeris data, but has the option of numerically integrating the orbit of Bennu, simultaneously with that of OSIRIS-REx. In this case, relevant force models are applied to Bennu. From work related to Mercury \cite{genova2016a}, we find that the inclusion of major asteroids in the force integration is important to accurately reproduce the JPL ephemerides through numerical integration. As such, we plan on accounting for the largest asteroid perturbations on Bennu during the OSIRIS-REx mission. Of the 343 asteroids included in DE432, we find that the bulk of these gravity accelerations will be from (1) Ceres, (4) Vesta, and (2) Pallas, with additional contribution from (15) Eunomia and (3) Juno (Figure~\ref{fig6}). GEODYN can also model and estimate tidal perturbations, through parameters such as the Love number k$_{2}$. While important for studies of Mars \cite{genova2016b}, the Moon \cite{lemoine2014}, and Mercury \cite{mazarico2014}, Bennu has too small of a radius for significant tides.

	A number of non-conservative forces are typically considered in orbit determination. Due to the low mass of Bennu, solar radiation pressure is especially important. The OSIRIS-REx spacecraft is modeled as a series of simple panels with assigned (and estimable) diffuse and specular reflectivity coefficients, following the macro-model approach of Marshall and Luthcke (1994). Improved fidelity will be achieved by modeling the self-shadowing between the plates \cite{mazarico2009}. During our recent work on the GRAIL mission \cite{lemoine2014}, we also developed a raytracing model to compute the solar flux received at the spacecraft in orbit around an airless body, accounting for actual shape and solar limb darkening. Given the terminator orbit configuration of OSIRIS-REx, no eclipse is expected in the early mission phases, but we will use this new capability if necessary. Planetary radiation accelerations due to short-wavelength radiation (`albedo') reflected and long-wavelength radiation (`thermal') emitted by Bennu are routinely considered in planetary work, and the near-spherical expected shape of Bennu allows the use of the same models \cite{lemoine1992}. Recent, independent developments to better deal with irregular shapes (e.g., 67P/Churyumov-Gerasimenko) could be leveraged if required.

	Spacecraft maneuvers adversely affect orbit reconstruction quality, and are thus typically avoided in gravity recovery analyses \cite{lemoine2014,mazarico2014}. Small thrusting events are however important to model and reconstruct in the case of the OSIRIS-REx mission as they are a critical, enabling the low-altitude reconnaissance passes and ultimately the TAG sampling. As such, we will use GEODYN's capability to model and estimate instantaneous 3-D velocity changes.

	To achieve the best orbit reconstruction quality, it is sometimes necessary to include a set of empirical accelerations in the estimation process, to account for unmodeled, systematic errors. GEODYN can model constant and once-per-revolution accelerations in the along-track, cross-track and radial directions of arbitrary duration, and estimate them iteratively. This was for example successfully used during LRO and GRAIL data analysis with very different strategies: \cite{mazarico2012} used a single along-track acceleration per 2.5-day arc, while \cite{lemoine2013} estimated sets of 6 acceleration parameters every 30 minutes for each GRAIL spacecraft. Of course, a smaller number of empirical accelerations is desirable if predictions are to be computed.

	Other force models were developed for high-precision analysis of Earth geodetic data but are often ignored in planetary work because of their small magnitude compared to other error sources. In the case of OSIRIS-REx, these are more significant because of Bennu's small mass and may thus need to be considered during proximity operations. Forces due to thermal anisotropy of the spacecraft and due to antenna radiation recoil have been modeled for the TOPEX-Poseidon mission \cite{marshall1992}. For OSIRIS-REx, we will use spacecraft telemetry of temperature sensors and modeling external to GEODYN to build time series of spacecraft-fixed acceleration perturbations that GEODYN will during integration.

	As described in Section~\ref{sec:orientationrepresentation}, the orientation of Bennu can be modeled and estimated using either an analytical approach or a dynamical numerical model. The orientation is important as it drives the computation of gravity field accelerations and planetary thermal radiation. In support for the OSIRIS-REx mission, the dynamical orientation model described above was implemented and tested, including the ability to estimate any or all of the twelve orientation parameters, through the computation of up to 144 variational equations (see Appendix, Section~\ref{sec:appendix}).

During orbit integration, GEODYN uses measurement models to compute the expected values of the considered observables. Reproducing the measurement geometry at appropriate times, GEODYN computes the expected values based on the current trajectory and model parameters. Many measurement types are implemented in GEODYN, which enables such modeling in a wide variety of geometries. In addition to geometrical configuration, observables rely on measurement corrections, that for example account for the effects of tropospheric delays due to particular conditions (temperature, pressure, water vapor partial pressure) at the time of measurements. We refer the reader to \cite{lemoine1997} for a detailed discussion of measurement correction models associated with atmospheric, tidal movement, station position, and relativistic effects.

\subsubsection{Radio Measurements}
\label{sec:radio}

The primary data used in planetary geodesy studies are the radio tracking Doppler and Range measurements acquired by the ground-based Deep Space Network (DSN, \cite{thornton2003}) using the spacecraft telecommunication system. Given the obvious need to retrieve telemetry and scientific data to make a planetary mission successful, relatively little hardware modifications are necessary to enable a high-quality geodetic investigation, and such investigations are typically part of flown planetary missions.

The radio wavelength used to communicate is important due to its associated intrinsic noise level and the expected noise level due to solar plasma (currently not correctable due to their stochastic nature and lack of data to precisely model the solar corona). While early spacecraft hardware operated in the S-band ($\sim$2.2GHz), recent planetary missions have operated in the X-band ($\sim$8GHz) which also provides a benefit in terms of data bandwidth. OSIRIS-REx operates in the X-band. In the future, tracking at Ka-band ($\sim$32GHz) is expected to further enhance the suite of possible scientific objectives (e.g., \cite{iess2009,hensley2016}).

	GEODYN can analyze unramped and ramped Range and Doppler radiometric data, and this capability has been demonstrated and exercised numerous times \cite{mazarico2012,mazarico2014,lemoine1997,lemoine2001,smith2012}. Tropospheric and ionospheric corrections can be applied using a variety of models and zenith mapping functions.

\subsubsection{Altimetry Measurements}
\label{sec:altimeter}

The GEODYN II software was essential to the success of the altimetric investigations by the Mars Orbiter Laser Altimeter (MOLA; \cite{smith1999}) and by the Lunar Orbiter Laser Altimeter (LOLA; \cite{smith2010,smith2016}), as the MGS and LRO trajectories were reconstructed with GEODYN. In addition, the altimetric ranges obtained by these instruments were folded in the orbit determination process, in the form of altimetric crossovers. Knowing that the planetary radius at groundtrack intersections did not change (except for small deformations due to tides), these measurements were used as constraints on the spacecraft orbit and altimeter pointing.  While this canonical crossover measurement type is well-adapted to large planetary bodies and narrow-groundtrack altimeters (it was also used by \cite{smith2012} on MESSENGER to validate early gravity field solutions), the OSIRIS-REx laser altimeter operates sufficiently differently (e.g. raster scanning at close proximity) that it is not practical at Bennu. Instead, we intend to use the OLA altimetry data in two different ways.

	Given prior information on planetary shape, the range to the surface as measured by OLA can be directly compared to a prediction using that shape and the spacecraft position and attitude. This simple measurement, called `direct altimetry', was key to calibrate the pointing of the GLAS instrument onboard ICESat. \cite{luthcke2000} proposed and leveraged large off-nadir slews while flying over oceans to estimate pointing parameters, which were then used to improve the geolocation of returns over the polar regions. The topography of the oceans was relatively well-known as it follows the geoid, and small errors, due to waves for example, were found not to affect the ability of the GLAS data nor contribute to the pointing recovery. We can extend this technique to short-scale, high-resolution targets. Bennu's shape will be known from image-based and from OLA altimetry models built from multiple acquisitions. We can use these shape models as a basemap to match individual altimetric tracks during orbit determination, adjusting both spacecraft position and pointing to minimize the discrepancies.

     In section~\ref{sec:pointing} we will discuss in more detail the estimation of pointing parameters for altimetry. For altimetry it is necessary to provide two estimated pointing correction angles: one in roll and the other in pitch. Camera data also require pointing correction angles and require an additional angle in yaw. As it turns out, the OLA scan naturally acquires data over a range of pointing angles that are phased in roll and pitch, much like the designed calibration maneuvers of [19]. Without a mechanism such as OLA's scanning mirror, the calibration maneuvers are typically spacecraft activities designed to acquire data over different pointing conditions so that position and pointing errors can be distinguished. We note that calibration maneuvers are not required for the camera data because many landmarks are visible in each image and in effect provide enough attitude information through parallax.
     	
	To take full advantage of the two-dimensional nature of the OLA altimetric raster scans, we will also use a crossover-like measurement recently implemented in GEODYN. Described in \cite{mazarico2010}, the so-called `swath crossovers' rely on the prior adjustment of two scans to each other. This relative positioning provides strong 3-D constraints between geolocated Cartesian coordinates that can be passed on to GEODYN to be replicated during orbit determination. One key aspect of this differenced measurement type compared to the direct altimetry is that it does not require nor depend on an a priori shape, and is insensitive to absolute pointing errors.

\subsubsection{Image Measurements}
\label{sec:image}

In preparation for the LRO and Dawn missions, a capability to process image data was implemented in GEODYN.

The traditional image landmark observables (e.g., \cite{konopliv2015}) are now available in GEODYN. Surface features need to first be recognized and matched across multiple, overlapping images. Their Bennu-fixed coordinates are typically obtained externally (through stereo-photoclinometry, SPC, or stereo-photogrammetry, SPG) and used as a priori. The pixel locations of these landmarks within the image collection are the geodetic data to be analyzed in GEODYN. A camera model is required to translate from vectors from landmark to spacecraft in the spacecraft-frame Cartesian space into camera pixel-space. The landmark positions, camera model parameters (e.g., focal length, distortion), and camera pointing biases can be estimated simultaneously by GEODYN, along with all other normal model parameters. This `image landmark' measurement is an absolute measurement, as it leverages the observation at one epoch to constrain spacecraft position and pointing, given landmark position and Bennu orientation.

	In contrast, another image-based measurement type developed by our group \cite{centinello2015} is the `image crossover' measurement. Akin to altimetric crossovers (e.g., \cite{rowlands1999}), it is a differenced measurement which in essence dictates that vectors pointing from the spacecraft through identified camera pixels should intersect in the Bennu-fixed frame. The camera model is now used to translate the pixel coordinates into a vector in physical space (the opposite of what is done for landmarks). Intuitively simple, this geodetic constraint is in some sense weaker than the previous landmark measurement, because a larger number of possible spacecraft trajectory and pointing solutions allow the rays to intersect. However, it does not rely on any prior information about the landmark position, and is insensitive to common-mode pointing biases. The link between spacecraft states at distant times can also help the solution, in particular of the orientation because well-behaved orbits can benefit other time periods with otherwise poorer knowledge. Finally, the creation of those measurements simply requires image processing tools to match feature pairs, which can be rapidly applied to new sets of images, in contrast to requiring the more involved SPC or SPG processing to create landmark data.

\subsubsection{Simulation Capabilities}
\label{sec:simulation}

While GEODYN is primarily designed to be used to process tracking data acquired by active flight missions, it provides the option to simulate all available measurement types. Comprehensive full simulations can be built, using the same measurement and force modeling capabilities. Once the artificial data are created and noise added, realistic errors and perturbations can be imposed on the initial state and a priori parameter values. GEODYN is then used to perform the normal orbit determination process, and both the actual errors in the recovered values and their formal uncertainties can give insight into the quality of the recovery and the potential presence of systematic errors. Of the many simulation efforts with GEODYN, several analyzed altimetry data to recover orbital and geophysical parameters. \cite{wahr2006} investigated the use of altimetric crossovers in orbit around Europa to recover tidal parameters (the Love numbers h$_{2}$ and k$_{2}$) to constrain the ice shell thickness. \cite{rowlands2009} presented simulation results using multi-beam crossovers enabled by the LOLA five-beam pattern. \cite{mazarico2015} revisited the subject of tides at Europa, using the data acquired by a long-range altimeter during a multiple-flyby mission. The OSIRIS-REx mission, and future mission concepts, can benefit from the detailed simulations enabled by GEODYN, as will be demonstrated in Section~\ref{sec:wobblerecovery}.

\subsection{Observation Types and Associated Measurement Partial Derivatives for Orientation Recovery}

\label{sec:obstypespartials}

\subsubsection{Observation Types}
\label{sec:obstypes}

Although all observations are tied to Bennu's orientation, the strength of the various observation types described above varies. Bennu's orientation is a weak factor in the force model of the OSIRIS-REx spacecraft. Indeed, although the OSIRIS-REx trajectory is sensitive to gravity and although the orientation of the gravity field is tied to Bennu's orientation, the orientation-trajectory link is weak because Bennu's gravity field is weak. This orientation-trajectory link affects all data types, but some measurements are vastly more sensitive than others to the orientation.
In addition to the limitations imposed by weak gravity, this link is only an indirect effect in that it does not provide information about the orientation of Bennu at any particular epoch. The orientation-trajectory link is an integrated effect. Only measurements that observe surface features can provide an an additional direct or geometric effect. Measurements with the direct effect can provide a detailed time history of orientation at individual epochs.

An interesting illustration of this fact can be drawn from our own experience during a verification and validation simulation effort for OSIRIS-REx (see Section~\ref{sec:setup}). The artificial observations for radiometric and landmark data were created in separate runs. By mistake, significantly different orientation models were used. When reducing the landmark data, using the wrong orientation model produce very large residuals, indicating a clear mismatch between trajectory and measurement modeling. However, this was barely noticeable in the Doppler data residuals, because the trajectory changes introduced by integrating the gravitational accelerations computed from the incorrect asteroid frame were so small and did not lead to large deviations from the true, self-consistent trajectory.

Camera and altimeter measurements have the direct geometric effect described above as well as the indirect trajectory effect. These measurement types directly observe surface features. In terms of camera-derived observations, a landmark measurement is thus very sensitive to the instantaneous $\alpha$, $\delta$, and W. Despite an equally strong geometric effect, landmark crossovers, because of their differential nature, are not sensitive to instantaneous values, but rather to orientation changes between the two measurement epochs of the crossover pair.
Around small bodies, the geometry of the observation pairs is typically quite different, so in practice the landmark crossovers are sensitive to the instantaneous orientation at both image epochs.
This situation is mirrored for the altimetric data: direct altimetry is sensitive to instantaneous orientation, while altimetric crossovers are mostly sensitive to relative changes in orientation. We note here, in anticipation of Section~\ref{sec:fullsim}, that undifferenced  measurements are critical to estimate the pointing of the camera and altimeter. The differenced (crossover) measurements are not sensitive to absolute pointing, only relative pointing changes. In order to estimate Bennu's orientation while simultaneously estimating instrument pointing biases, a combination of absolute and differential measurements is helpful.

The radio tracking data, acquired from Earth-based stations with no direct tie to Bennu's surface, have only the indirect trajectory effect. However, the DSN data are still important to conduct orbit determination of OSIRIS-REx around Bennu. Indeed, this lack of geometric effect is valuable for orbit stabilization in the inertial frame. The camera and altimeter data have no link to the inertial frame. Without the DSN data it would be difficult if not impossible to get the correct absolute orientation of Bennu.

\subsubsection{Associated Measurement Partial Derivatives}
\label{subsec:assmeaspds}

As was discussed in the previous section, all tracking measurement types associated with OSIRIS-REx are sensitive to Bennu's orientation through the indirect link provided by gravity into the trajectory of the OSIRIS-REx spacecraft. In other words, the 12 initial state parameters of Bennu orientation are force model parameters of the OSIRIS-Rex trajectory. As force model parameters, the partial derivatives of the spacecraft position with respect to the 12 orientation parameters are numerically integrated along with other force model parameters (like gravity coefficients) in the OSIRIS-REx variational equations. The variational equations for the OSIRIS-REx spacecraft require that the explicit partial derivatives of spacecraft acceleration with respect to each force model parameter be computed at each integration step. In the case of the 12 Bennu orientation parameters, this requires output from the separate set of variational equations that have been added to GEODYN for the orientation numerical integration. The inclusion of the Bennu orientation parameters as spacecraft force model parameters required a double set of linked variational equations.     

For each tracking measurement type, the measurement partials with respect to the orientation parameters has a component coming from the partial derivative of the measurement with respect to the spacecraft position chained together with the partial (discussed just above) of the spacecraft position with respect to the orientation parameters. For DSN data this is the only component. For camera and altimeter data there is a second and stronger direct effect component of the measurement partial derivative with respect to orientation parameters. For this second component, we compute the partial derivative of the measurement with respect to the $\alpha$, $\delta$ and W angles of Bennu at the measurement epoch. This is in turn chained with the partial derivatives of the three angles orientation angles with respect to the initial state parameters (from the orientation variational equations). For camera and altimeter measurements the two components are added together. 

\section{Simulation of Wobble Recovery}
\label{sec:wobblerecovery}

\subsection{Sensitivity and Linearity of the Orientation Parameters}
\label{sec:sensitivity}

As noted in Section~\ref{sec:practicalwobble}, one complication of using the dynamical approach is that the time history of orientation is highly non-linear with respect to the 6 moment of inertia parameters.

We performed tests estimating the initial state from the truth time series when various moments are in error. We first generated a `truth orientation' time history (series of $\alpha$, $\delta$, and W) for 7 days based on the a priori parameters shown in Eq.~\ref{eq2}.
We then estimated the moments of inertia from these angles, starting from perturbed moment values but a correct initial orientation state. As before, this is an optimistic case given that no OD filter noise is present in the estimation and that the measurements are `perfect' (direct observations of the orientation angles).

We find that if the starting off-diagonal moments for Bennu are `order of magnitude' correct, the wobble recovery is relatively insensitive to off-diagonal moment errors. If the error in the diagonal moments is less than $\sim$3\%, the orientation wobble state can be recovered. From Fig.~\ref{fig2}c, it appears the current uncertainties in shape are sufficiently small to fulfill that criterion.

In reality, the initial state will not be known, and we will need to estimate the initial state and the diagonal moments simultaneously. In a gravitational torque-free case, which is the case at Bennu, the three diagonal moments, $I_{xx}$, $I_{yy}$, and $I_{zz}$, cannot be estimated independently, and we need to add a constraint, such as fixing $I_{xx}^2$ + $I_{yy}^2$ + $I_{zz}^2$ to an a priori value (which is equivalent to constraining the total mass). Tests show that the $I_{xx}$, $I_{yy}$, and $I_{zz}$ parameters are highly non-linear and difficult to estimate robustly when the estimation period lengthens. Solutions much longer than three days decay due to the non-linearity of $I_{xx}$, $I_{yy}$, and $I_{zz}$. Longer solutions are only possible if these parameters are known and held fixed.

\subsection{Simulation Setup}
\label{sec:setup}

The simulation was based on the Orbit Determination Thread Test 3 (ODTT3) simulation effort conducted internally by the OSIRIS-REx Flight Dynamics Team in early 2015. It focused on the Orbital Phase B mission phase. Extensive details on the assumptions, methodology, and results are presented in \cite{getzandanner2016}. Here, we adopted the general setup, such as orbital configuration, observation geometry, and data coverage, but we re-simulated the observations with different assumptions, in particular of the orientation history of Bennu. In the 4-day analysis period, the first three days contain various radio tracking passes, image acquisitions and altimetry scans. The spacecraft trajectory is integrated forward over the last day to allow orbit prediction quality assessment. Of note, the arc contains 42 images and 42 OLA scans. Those scans follow the planned raster scan pattern for Orbital Phase B, but were downsampled significantly (by a factor of 25) for computational reasons. After three days of data collection, differenced measurements are available thanks to overlaps between images and between OLA scans.

\subsection{Separability of Orientation and Instrument Pointing Errors}
\label{sec:pointing}

In order to model the camera and altimeter observations and use them during orbit determination, pointing knowledge to relate instrument reference frame to spacecraft body reference frame to planet-fixed reference frame is required. The high-resolution instruments onboard OSIRIS-REx will require accurate pointing knowledge to contribute to the geodetic estimation, beyond what is possible to achieve with onboard sensors. That is in part because of the lack of an optical bench on OSIRIS-REx, meaning that the transformation to inertial provided by the star trackers cannot be rigidly tied to the instrument orientation. As such, we anticipate pointing errors would need to be estimated at each observation epoch. Can the data support this approach? Multiple landmark observations are expected to be available within each camera image, which should be sufficient for epoch-by-epoch estimation.
For the purpose of pointing correction, we consider that an entire OLA 150-second scan can be corrected by a single pair of pointing parameters.
The OLA scans provide a strong geometry which allows scan-by-scan pointing estimates. Within each 150-s scan, the OLA pointing is stable to 30 arcseconds, which is tolerable in the 1-km radius orbit ($\sim$10 cm).

Nevertheless, one difficulty that comes with estimating pointing parameters at every observation epoch is that these camera and altimeter parameters are highly correlated with the Bennu orientation parameters.

We performed a small simulation to test the separability of epoch by epoch pointing parameter and orientation parameters.
We modeled the asteroid orientation with the analytical model using periodic terms with an amplitude of one degree in both $\alpha$ and $\delta$. Although these amplitudes do not correspond to a wobble, they suffice to check the separability of orientation and attitude.
We simulated landmark and altimeter range observations over 4.5 days, based on the ODTT3 setup described above. Our truth model includes periodic variations in $\alpha$, $\delta$, W to produce a 1$^{\circ}$ `wobble'. We first simulated the observations: 293 landmarks over 42 images; 258,871 altimetric range observations, spread over 42 OLA scans (150 seconds each; the anticipated data was decimated by a factor of 25). To understand the intrinsic weakness of estimating either or both of the pointing and orientation parameters, the observations are free of noise and of initial pointing error, and we used the simulated data to test orientation recovery under perfect conditions, with no force or measurement model errors. All parameters start at their truth values, except for the $\alpha$, $\delta$ and W periodic terms (that can describe wobble-like dynamics) which are set to zero initially. The initial residual RMS, without the periodic terms, is 53 cm for the direct altimetry data, and 7.7 pixels for the image landmarks. We conducted separate simulations with either altimetric or landmark data, and our simulation results are very similar for each measurement type.

In a first test, we estimated pointing parameters at the 42 epochs, despite the fact that no pointing error were present. We adjusted either three angles for the camera observations (roll, pitch, and yaw) or two angles for the altimeter observations (roll and pitch). Such an inversion reduces the altimeter range RMS from 53 cm to 23 cm, and the landmark data RMS from 7.7 pixels down to 1.5 pixels. This means that instrument pointing parameters are highly correlated with orientation parameters, and can account for much of the signal that a wobble would create.

In our second test, we held the instrument pointing parameters to the values derived above, and performed a second inversion of the orientation periodic terms alone. The recovered `wobble' amplitude was greatly reduced compared to the full, true amplitude. This clearly demonstrates that one should not estimate instrument pointing and orientation parameters separately (sequentially), because of their high correlation.

In our third test, starting from no periodic terms (zero a priori), we estimated both the orientation periodic terms and the instrument pointing biases simultaneously. We find that in this case, the orientation periodic terms can be estimated almost perfectly. Although the recovery was near perfect, we emphasize that there were no systematic or random errors in this simulation. But this shows that in principle, epoch-by-epoch pointing parameter estimation is possible and not detrimental to the recovery of body orientation.

\subsection{Full Simulation with Systematic Errors}
\label{sec:fullsim}

The simulations described above had no error sources (only the co-estimation of highly correlated parameters). Here, we perform a full simulation with systematic errors to represent a realistic situation for OSIRIS-REx during Orbital Phase B. 

This simulation scenario is also based on ODTT3, but with new simulated data. We created artificial data over a span of 4.5 days, based on a set of truth trajectory and models. The 1-degree wobble discussed in Section 3.3 was considered. All measurement types described in Section~\ref{sec:geodyn} that we anticipate to be collected during Orbital Phase B were included, at a realistic cadence allowing for DSN tracking periods and camera and altimeter data acquisition periods. The resulting dataset includes: 816 landmark observations (at 42 epochs as previously); 731 landmark crossovers; 1914 DSN Doppler observations; 140 DSN Range observations over two passes; 275,562 altimeter ranges collected over 42 OLA scans; 27 altimeter patch crossovers obtained from overlapping OLA scan pairs, each with independent horizontal and radial constraints.

	We inserted systematic errors into this simulation, similar to those used in ODDT3 \cite{getzandanner2016}. Several gravity coefficients are set significantly away from truth.
Panel reflectivities (related to solar radiation) are also perturbed. A 1 m-level Gaussian error was added to the landmark coordinates. The OLA scans were given periodic roll and pitch errors (with a half-scan period), with an amplitude of 30 arcseconds.

	These errors make for a realistic simulation with large systematic effects, far from an optimistic case. To illustrate this, we computed the residuals obtained from an initial run with only these systematic errors, all other parameters being set to truth (including initial spacecraft state). We then performed an OD solution, only letting the initial spacecraft state adjust. The results, shown in Table~\ref{tabnew}, show that the initial state itself can account for a large part of the large initial residuals, but that large systematic signatures remain, particularly for some measurement types (e.g., landmarks and altimetric constraints). This also shows why the use of a diversity of measurements is helpful to avoid systematic effects from affecting the estimation.

\begin{table}
\caption{Effect of systematic errors introduced to the gravity field coefficients, landmark coordinates, and altimeter pointing. The pre-fit column indicates the residual RMS when all other parameters are set to true values. The post-fit column shows the improvements due to the sole adjustment of the spacecraft initial state.}
\label{tabnew}
\centering\begin{tabular}{lllll}
\hline\noalign{\smallskip}
Type & Number of & Pre-fit  & Post-fit RMS & Unit \\
& Observations & RMS & & \\
\hline\noalign{\smallskip}
Landmark & 816 & 37.4 & 1.71 & pixel \\
\hline\noalign{\smallskip}
Landmark Crossover & 731 & 13.9 & 0.54 & m \\
\hline\noalign{\smallskip}
DSN Range & 140 & 52.5 & 1.92 & RU \\
\hline\noalign{\smallskip}
DSN Doppler & 1914 & 0.046  & 0.10 & Hz \\
\hline\noalign{\smallskip}
Altimeter Range & 275562 & 4.78 & 0.00 & m\\
\hline\noalign{\smallskip}
Altimeter Crossover & 27 & 21.1 & 11.48 & m \\
(Horizontal) & & & & \\
\hline\noalign{\smallskip}
Altimeter Crossover & 27 & 4.1 & 2.70 & m \\
(Vertical) & & & & \\
\hline\noalign{\smallskip}
\end{tabular}
\end{table}

	We then checked the effect of these systematic errors on the data, by starting with the true Bennu orientation values, and performing an orbit solution during which we estimated various parameters: initial state, solar radiation scale factor, pointing error parameters, and range biases per pass. Table~\ref{tab3} shows the pre- and post-fit residual RMS values, and indicates the level to which the orbit state adjustment can accommodate the systematic errors arising from holding certain parameters to values different from the truth.

\begin{table}
\caption{Residual RMS of the various measurement types used in the simulation, without any error on the orientation of Bennu. The residuals converge to near-noise level after orbit determination is performed.}
\label{tab3}
\centering\begin{tabular}{llll}
\hline\noalign{\smallskip}
Type & Pre-fit RMS & Post-fit RMS & Unit \\
\hline\noalign{\smallskip}
Landmark & 37.4 & 4.8 & pixel \\
\hline\noalign{\smallskip}
Landmark Crossover & 13.9 & 0.5 & m \\
\hline\noalign{\smallskip}
DSN Range & 52.5 & 1.9 & Range Unit\cite{DSN810005} \\
\hline\noalign{\smallskip}
DSN Doppler & 0.046 & 0.003 & Hz \\
\hline\noalign{\smallskip}
Altimeter Range & 4.78 & 0.69 & m \\
\hline\noalign{\smallskip}
Altimeter Crossover (Horizontal) & 21.16 & 2.87 & m \\
\hline\noalign{\smallskip}
Altimeter Crossover (Vertical) & 4.13 & 0.67 & m \\
\hline\noalign{\smallskip}
\end{tabular}
\end{table}

	We also imposed errors on the initial orientation state: $\dot{\alpha}_{0}$ and $\dot{\delta}_{0}$ were set to zero (no a priori wobble), and we imposed an initial error on $I_{yy}$ of 3\%. We estimated all of the normal arc parameters as above, plus 9 Bennu orientation parameters (initial state orientation and the three diagonal moments $I_{xx}$, $I_{yy}$, $I_{zz}$). Because we estimate the moments, the data span needed to be short, and we chose a length of 3 days. The number of available measurements is reduced from those originally simulated. Differenced measurements can reduce substantially, because they intrinsically require a long timespan to occur (and their occurrence tends to grow quadratically with time). In particular, only six OLA patch crossovers are available within a 3-day period, compared to 27 over the full 4.5-day span. A 2-day arc duration would be more desirable due to non-linearity issues, but arcs shorter than 3 days would not contain any OLA patch crossover, an important asset.

	Table~\ref{tab4} and ~\ref{tab5} present the results of two simulations, identical except for the weighting of the landmark data. Observation fits are generally good, which indicates orbit convergence. Many geodetic parameters are recovered well. For instance, the three diagonal moments are estimated with an accuracy of $\sim$1\%. As in Section~\ref{sec:practicalwobble}, we transform these angle errors into geolocation errors on the ground, which is most relevant to the TAG accuracy requirement for OSIRIS-REx. Table~\ref{tab6} shows that geolocation errors are nearly always smaller than 1 m in the 3-day reconstruction period. The method of computing geolocation errors will be discussed in the next section. The prediction performance was computed over the 1.5 days following data cut-off. In both landmark weight cases, it remains below 0.9 m in the RMS sense in the equatorial region (Table~\ref{tab6}), which is on the same order as the TAG accuracy requirement (1 m, \cite{berry2013}). A large Bennu wobble would thus contribute a notable part of the error budget. In the quiet Bennu case, without any large wobble, the maximum geolocation error can remain below 1 m RMS globally when landmarks are upweighted.
However, we note that not considering a wobble would yield much poorer results. Indeed, reproducing the simulation using the analytical approach without periodic terms, we obtain geolocation errors of $\sim4$ m and $\sim 5m$ RMS in the reconstruction and prediction periods, respectively.

\begin{table}
\caption{Summary of the tracking data used in the simulation. All available measurement types are used. Two cases for the weight given to the landmarks are shown. The downweighting of the landmark data does not significantly affect their level of fit, but improves the consistency, with the differential measurements in particular (landmark and altimeter crossovers).}
\label{tab4}
\centering\begin{tabular}{llllll}
\hline\noalign{\smallskip}
Type & Number of & Pre-fit  & Post-fit RMS & Post-fit RMS & Unit \\
& Observations & RMS & (downweighted & (upweighted & \\
& & & landmarks) &  landmarks) & \\
\hline\noalign{\smallskip}
Landmark & 555 & 282.2 & 4.9 & 4.8 & pixel \\
\hline\noalign{\smallskip}
Landmark Crossover & 472 & 50.2 & 0.3 & 0.4 & m \\
\hline\noalign{\smallskip}
DSN Range & 70 & 41.3 & 1.2 & 3.8 & RU \\
\hline\noalign{\smallskip}
DSN Doppler & 1064 & 0.029  & 0.002 & 0.006 & Hz \\
\hline\noalign{\smallskip}
Altimeter Range & 190269 & 6.64 & 0.60 & 0.85 & m\\
\hline\noalign{\smallskip}
Altimeter Crossover & 6 & 37.0 & 1.56 & 2.16 & m \\
(Horizontal) & &  &  &  &  \\
\hline\noalign{\smallskip}
Altimeter Crossover & 6 & 2.68 & 0.27 & 0.51 & m \\
(Vertical) & &  &  &  &  \\
\hline\noalign{\smallskip}
\end{tabular}
\end{table}

\begin{table}
\caption{Results of the estimation of the diagonal moments of inertia in two simulation cases.}
\label{tab5}
\centering\begin{tabular}{|l|l|l|l|lll|lll|}
\hline
Parameter & Unit & Truth & a priori & \multicolumn{3}{c|}{Downweighted landmarks} & \multicolumn{3}{c|}{Upweighted landmarks} \\
 & & ($\times 10^{15}$) & & Recovered & Error & Error (\%) & Recovered & Error & Error (\%) \\
\hline
$I_{xx}$ & kg.m$^{2}$ & 1.752 & truth & 1.7495 & 0.003 & 0.17 & 1.7700 & 0.018 & 1.03 \\
\hline
$I_{yy}$ & kg.m$^{2}$ & 1.820 & 1.872 & 1.8518 & 0.032 & 1.75 & 1.8363 & 0.016 & 0.88 \\
\hline
$I_{zz}$ & kg.m$^{2}$ & 1.968 & truth & 1.9897 & 0.022 & 1.12 & 1.9858 & 0.018 & 0.91 \\
\hline
\end{tabular}
\end{table}

\begin{table}
\caption{Simulated performance in geolocation after estimation of the orientation of Bennu, in the case of a large 1$^{\circ}$ wobble and in the case of a quiet Bennu. Results for each case are given for two different landmark data weights.}
\label{tab6}
\centering\begin{tabular}{|l|l|ll|ll|ll|}
\hline
 & Geolocation Errors (m) & \multicolumn{2}{c|}{$\lambda = 0-30^{\circ}$} & \multicolumn{2}{c|}{$\lambda = 30-60^{\circ}$} & \multicolumn{2}{c|}{$\lambda = 60-90^{\circ}$} \\
 & & RMS & Max. & RMS & Max.  & RMS & Max. \\
\hline
& & & & & & & \\
\multirow{15}{*}{\rotatebox{90}{1$^{\circ}$ wobble}}  & Initial (0-72 h) & 91.80 & 169.74 & 65.36 & 138.55 & 29.17 & 77.33 \\
& & & & & & & \\
& Initial (72-108 h) & 198.26 & 230.86 & 149.11 & 206.15 & 62.47 & 130.66 \\
& & & & & & & \\
& Solution (0-72 h) & 0.37 & 0.84 & 0.44 & 1.01 & 0.58 & 1.12 \\
& (downweighted) & & & & & & \\
& & & & & & & \\
& Prediction (72-108 h) & 0.89 & 1.89 & 1.29 & 1.96 & 1.51 & 2.33 \\
& (downweighted) & & & & & & \\
& & & & & & & \\
& Solution (0-72 h) & 0.33 & 0.61 & 0.42 & 0.76 & 0.51 & 1.00 \\
& (upweighted) & & & & & & \\
& & & & & & & \\
& Prediction (72-108 h) &0.85 & 1.70 & 1.16 & 1.98 & 1.37 & 1.98 \\
& (upweighted) & & & & & & \\
& & & & & & & \\
\hline
& & & & & & & \\
\multirow{12}{*}{\rotatebox{90}{quiet Bennu}}  & Solution (0-72 h) & 0.44 & 0.84 & 0.44 & 0.77 & 0.40 & 0.65\\
& (downweighted) & & & & & & \\
& & & & & & & \\
& Prediction (72-108 h) & 0.74 & 1.16 & 0.62 & 1.12 & 0.46 & 0.80 \\
& (downweighted) & & & & & & \\
& & & & & & & \\
& Solution (0-72 h) & 0.38 & 0.63 & 0.44 & 0.64 & 0.47 & 0.67 \\
& (upweighted) & & & & & & \\
& & & & & & & \\
& Prediction (72-108 h) & 0.70 & 0.98 & 0.59 & 0.89 & 0.43 & 0.73 \\
& (upweighted) & & & & & & \\
& & & & & & & \\
\hline
\end{tabular}
\end{table}

\subsection{Implications for Spin State Recovery and Geolocation Accuracy}
\label{sec:implications}

\begin{figure}
\centering\includegraphics[width=4.5in]{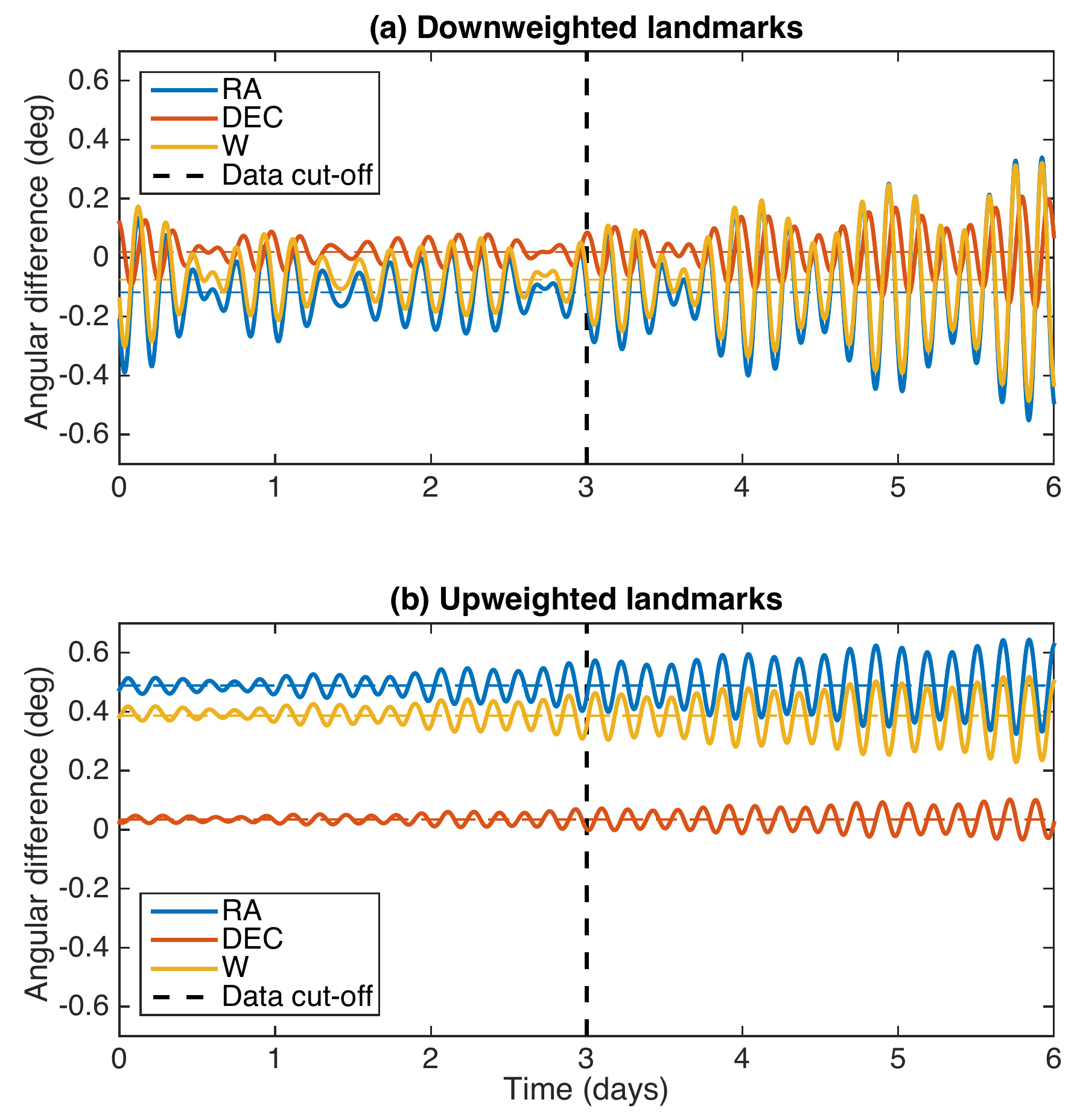}
\caption{Errors in the three inertial orientation angles ($\alpha$, $\delta$, W) after analysis of the radio tracking, altimetry, and image data, in our most comprehensive 1$^{\circ}$ wobble simulation cases. Results with different weighting for the landmark data are shown: (a) downweighted landmark data and (b) upweighted landmark data (close to noise level).The correlation between $\alpha$ and W leads to both being biased on average (thin dashed lines). Past the data cut-off date, the prediction performance degrades slowly. Downweighting the landmarks yields better average orientation angles. However, the geolocation performance is actually better with upweighted landmarks, thanks to orbit plane errors partially compensating the inertial pole position errors.}
\label{fig7}
\end{figure}

As indicated in Tables~\ref{tab4}-\ref{tab6}, we performed the simulation using two different weighting schemes for the landmark data.
In our original simulation, the landmark data were weighted close to their noise level (`upweighted' case), around 0.1 pixel. The data fits were good (Table 4), but when assessing the errors in orientation parameters, the time series (Figure~\ref{fig7}b) revealed relatively large biases for $\alpha$ and W compared to the truth values.
We performed an alternate simulation with lower weights for the landmark data (`downweighted' case). This downweighting, by a factor of 10 (to 1.0 pixel), significantly reduced the mean in the angular errors (Figure~\ref{fig7}a).
Each solution has advantageous characteristics that can be exploited depending on the application. The downweighted landmark solution has lower overall angular errors (good for astrometry), whereas the upweighted landmark solution has much smaller angular variations about the mean errors. It turns out that for the purposes of geolocation, mean errors are not as detrimental as departures from mean errors.
The better orientation recovery accuracy when downweighting the landmark data is an important finding, which we will use to inform our work during the actual OSIRIS-REx mission.

When computing geolocation errors from a set of spin state angular errors, it is tempting to simply rotate a set of body fixed coordinates on the surface of the asteroid to inertial coordinates. This would be done with the `truth' angles and then again using the solution angles that are in error. The two sets of inertial coordinates would be compared. However, this approach only considers spin state error and neglects the contribution of orbit error to geolocation error. 
During the OSIRIS-REx mission, we will be computing orbits using our estimated spin states. The spin state error will contribute to orbit error which will in turn have an effect on geolocation error. The geolocation errors reported in Table~\ref{tab6} were computed in the following manner. The `truth' orbit of the simulation was rotated to body fixed using the `truth' spin state. The latitude and longitude of each point along the trajectory was computed and then located on the asteroid surface. This represents the `truth' geolocation. The same tracking data  used to estimate the spin state was used together with the estimated spin state to make an orbit solution. The points along the new orbit solution were geolocate in the same way as the `truth' geolocation, except that the estimated spin state was used in the rotations. This provided the set of geolocated points in error that were compared to the `truth' geolocation.

While the astrometric performance is improved by downweighting the undifferenced measurements, we find that the geolocation performance actually degrades (Table~\ref{tab6}). Initially counter-intuitive in light of the previous results, this can be explained by the fact that the mean errors in orientation angles for the pole of Bennu are compensated in large part by similar, correlated errors in the absolute inertial orientation of the OSIRIS-REx orbit plane. When upweighted, the camera and altimetry undifferenced data will `drag' the orbit plane into an orientation that is consistent with the spin state (which will have larger spin state errors than when downweighted). For the purpose of minimizing the geolocation errors, especially important for executing the TAG maneuver and sampling, the landmark and direct altimetry data will thus provide a stronger tie to the Bennu-fixed frame.

Different combinations of the four camera and altimetric data types during the OSIRIS-REx mission will provide ways to leverage the scientific data and better achieve both astrometric accuracy and navigation goals.

In the orbit determination process, instrument pointing biases and Bennu orientation parameters are correlated, but the ability to use differenced and undifferenced measurements provides a powerful way to separate them. The undifferenced measurements are sensitive to absolute instrument pointing, while the differenced measurements cannot recover absolute pointing, but simply give constraints on the relative pointing differences. By downweighting the undifferenced measurements, they will not significantly contribute to the orientation parameters, but still allow good recovery of instrument pointing biases because they are sensitive to them and because the differenced measurements are less sensitive. The differenced measurement, by being relatively upweighted, will primarily determine the Bennu orientation. The reason we may want to prevent the undifferenced measurements from contributing to orientation is that the instrument pointing cannot necessarily be recovered to sufficient absolute quality, and thus that these types of measurements would necessarily degrade the orientation parameters referenced to the inertial frame, because of the high correlation. This is also true but to a lesser extent in the case of differenced measurements, so those provide the best way of determining Bennu orientation in inertial space, for astrometry purposes. Although this is clearly demonstrated for the camera measurement types by the above results, we expect this to also be the case for the altimetric measurement types.


\subsection{Why Is It So Hard to Get Geolocation to Better Than a Meter?}
\label{sec:whyhard}

The simulations presented above with both a large 1$^{\circ}$ wobble, and another conducted with a quiet Bennu (Table~\ref{tab6}), point to a final geolocation accuracy on the order of one meter. In the case where no wobble is present, the orientation can be recovered to slightly better than one meter, satisfying the orientation recovery requirement to achieve the larger TAG requirement.

There are three reasons why this value of one meter appears to be a minimum, floor value in the case of Bennu. These are outlined in the following sections.

\subsubsection{Small periodic variations caused by off-diagonal moments of inertia}
\label{sec:smallperiodic}

As shown in Section~\ref{sec:practicalwobble}, there is a limit to how quiet Bennu can be. That is because in practice, it will be nearly impossible to implement a perfectly principal axes body-fixed frame. Because of the off-diagonal moments of inertia, the closest that Bennu could come to a constant pole with a constant rotation rate translates to a position error just under a meter at the equator (Table~\ref{tab2}). 
Any simple analytical model with four parameters (constant $\alpha$ and $\delta$, W$_{0}$ and $\dot{\text{W}}$) will produce errors of this magnitude in the best of circumstances. The variations in orientation that are left unmodeled by the simple four parameter model can be picked up by adding periodic terms for $\alpha$, $\delta$ and W (two parameters in each angle, or six parameters total). The unmodeled variations in orientations can also be addressed by using a dynamic orientation model. Either approach comes with some drawbacks, which when combined with other error sources make it difficult to achieve ultra-precise results. The analytical periodic terms require prior information on the period of the variations, and any error in the period will cause problems. However, this period depends on the ratios of the moments of inertia, which are not perfectly known. These ratios must be determined from the dynamic orientation approach, but they are very non-linear parameters, requiring estimation from short arcs.

\subsubsection{Observation noise and force model errors}
\label{sec:obsnoise}

This simulation did not consider random observation errors (noise), so the results are necessarily optimistic. However, systematic errors are the largest contributor to the error budget. Several systematic errors can be considered.

	The errors in camera and altimeter instrument attitude are mostly recoverable by estimating epoch-by-epoch attitude parameters (for altimetry one epoch is a 150-second scan). The effect of these errors will be discussed below (Section~\ref{sec:instattitude}).

	The OLA attitude is also affected by short-wavelength errors. Each OLA scan spans 150 seconds and contains many thousands of individual altimeter ranges. During the 150 seconds of the scan, the attitude error oscillates about the mean attitude error, which has an amplitude of about 30 arcseconds and is estimated by 75s-period empirical (as above). The data were simulated without these short-wavelength errors on top of the long-wavelength effect. Because of their short-wavelength nature however, we can expect their average effect to be small and not affect the orientation solution beyond degrading slightly the quality of the OLA patch crossovers (e.g., noise on the constraints).

	Errors in landmark coordinates may also induce systematic effects. We considered such landmark errors in the simulation, and the simulated data were reduced using landmark coordinates to which 1-m RMS position Gaussian errors were imposed. However, if the errors are spatially correlated after stereo processing, additional systematic effects may result. 

Given the relatively large influence of solar radiation pressure on OSIRIS-REx in Bennu's weak gravity environment, errors in the force model, such as spacecraft shape and panel reflectivities, can affect the spacecraft trajectory. The nature of such errors (force model parameter bias) would naturally yield systematic errors. In our simulation, we accounted for such effects by simulating the data with one truth spacecraft model and performing the OD with another set with realistic errors (Section~\ref{sec:setup}).

Similarly, any error in the gravity field coefficients will produce systematic errors and biases in the reconstructed trajectory. Here again, following the ODTT3 simulation, we simulated the data with the truth gravity field (to degree and order 16) but the OD analysis was performed with large errors in the gravity coefficients.

To better understand the contribution of these systematics to geolocation error, we reran the `quiet Bennu' full simulation (Section~\ref{sec:fullsim}), but with no instrument pointing errors and no adjustment of any instrument pointing parameters. During the 3-day solution period, the global geolocation errors are $\sim$ 40 cm RMS (88 cm maximum), while over the 1.5-day prediction period, the errors increase slightly to $\sim$50-60 cm RMS (92 cm maximum). While smaller because of the reduced number of errors and adjustable parameters, these numbers are only slightly reduced, which shows that systematic errors have the most impact on geolocation error.

\subsubsection{Estimation of camera and altimeter attitude parameters}
\label{sec:instattitude}

The simulations described in Section~\ref{sec:pointing} showed that it is possible and almost certainly necessary in order to estimate orientation accurately to adjust three attitude parameters (roll, pitch and yaw) for each landmark epoch and two attitude parameters (roll and pitch) for each 150-second altimeter scan.
       The estimation of these attitude parameters does slightly degrade the orientation solution and the related geolocation. We performed a small simulation in the `quiet' Bennu case to determine how well the orientation parameters could be recovered from the simulated data. This recovery test had no errors on the data or on any of the modeled parameters. The orientation parameters were set at their truth values and allowed to adjust, along with the satellite state parameters. 
	The resulting geolocation errors after adjustment are small, around 10 cm RMS and always below 20 cm during the reconstruction period. In the 1.5 day prediction period, the geolocation does not degrade substantially ($\sim$ 10 cm RMS, and 23 cm maximum). This means that when no systematic errors are present, the geolocation errors are much smaller than the TAG requirement. Systematic errors in orientation, gravity, and force models are the most detrimental to the geolocation accuracy.

\section{Summary and Conclusions}
\label{sec:summary}

This simulation effort was designed to assess the implications of a wobble at Bennu for the OSIRIS-REx mission. We showed how a wobble would appear and affect the TAG accuracy in particular. We found that if the a priori off-diagonal moments of Bennu are order of magnitude correct, the wobble solution is relatively insensitive to those off-diagonal moment errors. During analysis, it may be necessary to estimate the $I_{xx}$, $I_{yy}$, and $I_{zz}$ moments, with constraints given the orientation is only sensitive to their ratio. Because these moment parameters are highly non-linear, arc lengths are best kept to 3 days or shorter. The estimation of epoch-by-epoch instrument pointing error parameters is possible when obtaining Bennu orientation solutions, but such parameters must be estimated simultaneously with orientation parameters to prevent large evaluation biases. The altimetry data are very helpful in separating orbit, instrument pointing errors, and Bennu orientation parameters. The combined use of differenced and undifferenced measurement types for both image and camera is also important to estimate the orientation parameters while needing to adjust the pointing of the instruments.  The undifferenced data (image landmark and direct altimetry) are most sensitive to absolute instrument pointing. The direct altimetry data (not the crossovers) need to be greatly down-weighted due to the sheer number of range observations (compared to other available data types) and because these range observations are sensitive to short-wavelength topography not necessarily captured in the available shape model. The differenced measurements (camera and altimeter crossovers) are especially useful to estimate orientation. They benefit from longer arc durations, because the number of overlapping images or altimetry scans typically increase quadratically with time. Too short an arc may also not contain any overlap. Our simulation indicated that orientation reconstruction and prediction accuracies of $\sim$1 meter can be obtained when using those data types.
We found that mean angular spin state errors are not necessarily detrimental to geolocation, thanks to their correlation with spacecraft orbit errors. For geolocation quality, reducing the departures from the mean is more important. Depending on the weighting strategy of difference and undifferenced data, solutions can be tailored for either geolocation performance or astrometric accuracy.
Even in the case of a large 1-degree wobble, the TAG accuracy requirement on geolocation error due to orientation knowledge should not pose a significant difficulty.

\begin{acknowledgements}
The authors acknowledge support from the NASA New Frontiers OSIRIS-REx project. This work was supported by NASA contract NNM10AA11C.
We thank two anonymous reviewers and the Editor-in-Chief for their comments which helped improve the manuscript.
The final publication is available at Springer via \url{http://dx.doi.org/10.1007/s00190-017-1014-1}.
\end{acknowledgements}

\section{Appendix}
\label{sec:appendix}

\subsection{Computing the planetary orientation parameters through dynamics} \label{appAcomp}

Clearly we need to be able to compute the orientation angles of Section~\ref{sec:orientation} at epoch $t$ in the dynamic case.  However, it is more natural to first compute related quantities at $t$ that are defined explicitly in the rotational equations of motion rather than the angles and their rates.  This essentially involves the numerical integration of the body-fixed axes in the J2000 frame.  This in turn involves the numerical integration of the angular velocities about each body-fixed axis.  From the axes we may compute the orientation angles at any time.  We may compute the initial state of the axes from the initial orientation angles.

With $\alpha$ the right ascension and $\delta$ the declination, the rotation from the body-fixed frame to the J2000 frame is given by
\begin{equation}
\mathbf{R}=
\left({
\begin{array}{ccc}
-S_{\alpha} C_W-C_{\alpha} S_{\delta} S_W & S_{\alpha} S_W-C_{\alpha} S_{\delta} C_W & C_{\alpha} C_{\delta} \\
C_{\alpha} C_W-S_{\alpha} S_{\delta} S_W & -C_{\alpha} S_W-S_{\alpha} S_{\delta} C_W & S_{\alpha} C_{\delta} \\
C_{\delta} S_W & C_{\delta} C_W & S_{\delta}
\end{array}}\right), \label{eq1ad}
\end{equation}
where $C_{[\cdot]}$ and $S_{[\cdot]}$ denote $\cos{(\cdot)}$ and $\sin{(\cdot)}$, respectively, of an angular argument.  If the cross-product between two generic vectors $\mathbf{u}$ and $\mathbf{v}$ is expressed as $\mathbf{u}\times\mathbf{v}=\mathbf{E}_{u}\mathbf{v}$, where
\begin{equation}
\mathbf{E}_{u}=
\left({
\begin{array}{ccc}
0 & -u_z & u_y \\
u_z & 0 & -u_x \\
-u_y & u_x & 0
\end{array}}\right), \label{eq1aad}
\end{equation}
is a traceless, skew-symmetric matrix and $\mathbf{u}=\left[{u_x\; u_y\; u_z}\right]^{\rm T}$, then this rotation is a function of time and can be interpreted in one of two ways:  (i) its orthogonal columns define the body-fixed axes of the rigid object in the J2000 frame such that the rate of change of this rotation is given by
\begin{equation}
\mathbf{\dot{R}}(t)=\mathbf{E}_{\tilde{\omega}}(t)\mathbf{R}(t), \label{eq2ad}
\end{equation}
where $\boldsymbol{\tilde{\omega}}(t)$ is the angular velocity of the object in the J2000 frame, or (ii) its orthogonal rows define the J2000 axes in the body-fixed frame of the rigid object such that the rate of change is given by
\begin{equation}
\mathbf{\dot{R}}^{\rm T}(t)=\mathbf{E}_{\omega}^{\rm T}(t)\mathbf{R}^{\rm T}(t), \label{eq3ad}
\end{equation}
where $\boldsymbol{\omega}(t)$ is the angular velocity of the object in the body-fixed frame.  Equation~\ref{eq3ad} leads to the following relationship
\begin{equation}
\boldsymbol{\omega}(t)=
\left({
\begin{array}{ccc}
C_{\delta} S_W & -C_W & 0 \\
C_{\delta} C_W & S_W & 0 \\
S_{\delta} & 0 & 1
\end{array}}\right)
\left({
\begin{array}{c}
\dot{\alpha}(t) \\
\dot{\delta}(t) \\
\dot{\text{W}}(t)
\end{array}}\right), \label{eq4ad}
\end{equation}

The Euler equations of motion for a rigid body with one point fixed \cite{goldstein1965} relate this angular velocity to its angular acceleration in the body-fixed frame under the affects of an applied torque, $\boldsymbol{\tilde{\tau}}(t)$, in the J2000 frame (here considered to be due only to the sun) such that
\begin{equation}
\boldsymbol{\dot{\omega}}(t)=\mathbf{I}_c^{-1}\left[{\mathbf{R}^{\rm T}(t) \boldsymbol{\tilde{\tau}}(t)-\boldsymbol{\omega}(t)\times\mathbf{I}_c \boldsymbol{\omega}(t)}\right], \label{eq5ad}
\end{equation}
where $\mathbf{I}_c$ is the moment of inertia matrix in the body-fixed cartesian frame.  Equations~\ref{eq3ad} and \ref{eq5ad} are the differential equations that we numerically integrate.  Although equation~\ref{eq3ad} is of primary interest, it cannot be integrated independently from equation~\ref{eq5ad}.  Furthermore, the numerical integration of equation~\ref{eq5ad} provides the vector corresponding to the instantaneous spin axis.  We may compute the initial conditions of the spin axis, $\boldsymbol{\omega}(t)$, from the initial orientation angles and their rates using equation~\ref{eq4ad}.

If we define $\boldsymbol{\rho}(t)=vec\left({\mathbf{R}^{\rm T}(t)}\right)$, where the $vec\left({\cdot}\right)$ operator stacks the columns of its matrix argument, then together, equations~\ref{eq3ad} and \ref{eq5ad} and initial conditions describe the time evolution of $12$ variables, i.e, the $3$ elements of $\boldsymbol{\omega}(t)$ and the $9$ elements of $\boldsymbol{\rho}(t)$, and can be represented by the augmented system
\begin{equation}
\boldsymbol{\dot{\zeta}}(t)=\mathbf{f}(t,\boldsymbol{\zeta}(t),\mathbf{s}),\quad\boldsymbol{\zeta}(t_0)=\boldsymbol{\zeta}_0 \label{eq6ad}
\end{equation}
where
\begin{equation}
\boldsymbol{\zeta}(t)=
\left({
\begin{array}{c}
\boldsymbol{\omega}(t) \\
\boldsymbol{\rho}(t)
\end{array}}\right), \label{eq7ad}
\end{equation}
and $\mathbf{s}$ is a vector of length $6$ containing the upper-triangular elements of the symmetric matrix $\mathbf{I}_c$, i.e., $\left({I_{xx}, I_{yy}, I_{zz}, I_{xy}, I_{xz}, I_{yz}}\right)$.  Thus, equation~\ref{eq6ad} may be integrated from an initial epoch $t_0$ to time $t$ using equations~\ref{eq1ad} and \ref{eq4ad} as a link between $\mathbf{R}(t)$, $\boldsymbol{\omega}(t)$, and the orientation angles and their rates.

In the case where $\mathbf{R}^{\rm T}(t){=}\left[{\mathbf{e}_x(t)\;\mathbf{e}_y(t)\; \mathbf{e}_z(t)}\right]$ represents the J2000 axes $\left({\mathbf{e}_x,\mathbf{e}_y,\mathbf{e}_z}\right)$ in the principal axis frame of the body, then $\mathbf{I}_c$ is diagonalized and equation~\ref{eq6ad} can be written in the more familiar form
\begin{equation}
\left({
\begin{array}{c}
\dot{\omega}_x(t) \\
\dot{\omega}_y(t) \\
\dot{\omega}_z(t) \\
\mathbf{\dot{e}}_x(t) \\
\mathbf{\dot{e}}_y(t) \\
\mathbf{\dot{e}}_z(t)
\end{array}}\right)=
\left({
\begin{array}{c}
\left[{\tau_x(t)-\left({I_{zz}-I_{yy}}\right) \omega_y(t) \omega_z(t)}\right]/I_{xx} \\
\left[{\tau_y(t)-\left({I_{xx}-I_{zz}}\right) \omega_x(t) \omega_z(t)}\right]/I_{yy} \\
\left[{\tau_z(t)-\left({I_{yy}-I_{xx}}\right) \omega_x(t) \omega_y(t)}\right]/I_{zz} \\
-\boldsymbol{\omega}(t)\times\mathbf{e}_x(t) \\
-\boldsymbol{\omega}(t)\times\mathbf{e}_y(t) \\
-\boldsymbol{\omega}(t)\times\mathbf{e}_z(t)
\end{array}}\right), \label{eq8ad}
\end{equation}
where $I_{xx}$, $I_{yy}$, and $I_{zz}$ are the diagonal elements of $\mathbf{I}_c$ and $\tau_x$, $\tau_y$, and $\tau_z$ are now the torques in the principal axis frame.

When working in the principal-axes frame, in the special case when $I_{zz}\ge I_{xx}=I_{yy}$ and the torque is negligible, equation~\ref{eq8ad} becomes
\begin{equation}
\left({
\begin{array}{c}
\dot{\omega}_x(t) \\
\dot{\omega}_y(t) \\
\dot{\omega}_z(t)
\end{array}}\right)=
\left({
\begin{array}{c}
-\beta \omega_y(t) \omega_z(t) \\
 \beta \omega_x(t) \omega_z(t) \\
0
\end{array}}\right), \label{eq13ad}
\end{equation}
where $\beta={\left({I_{zz}-I_{xx}}\right)\over I_{xx}}$,
from which it is clear that $\omega_z(t)$ is a constant, say $\omega_{z0}$.  This leads to
\begin{equation}
\left({
\begin{array}{c}
\dot{\omega}_x(t) \\
\dot{\omega}_y(t)
\end{array}}\right)=
\left({
\begin{array}{cc}
0 &-\beta \omega_{z0}\\
\beta \omega_{z0} & 0
\end{array}}\right)
\left({
\begin{array}{c}
\omega_x(t) \\
\omega_y(t)
\end{array}}\right). \label{eq14ad}
\end{equation}

The solution is
\begin{equation}
\left({
\begin{array}{c}
\omega_x(t) \\
\omega_y(t) \\
\omega_z(t)
\end{array}}\right)=
\left({
\begin{array}{ccc}
\cos(\beta \omega_{z0} t) &  ~~-\sin(\beta \omega_{z0} t) &~~ 0 \\
\sin(\beta \omega_{z0} t) &  ~~\cos(\beta \omega_{z0} t) &~~ 0 \\
0 & ~~0 &~~ 1
\end{array}}\right)
\left({
\begin{array}{c}
\omega_x(0) \\
\omega_y(0) \\
\omega_z(0)
\end{array}}\right), \label{eq15ad}
\end{equation}
and as a result $T=2\pi\beta^{-1}\omega_{z0}^{-1}$ is the period of the wobble.

\subsection{Estimating the planetary orientation parameters and moments of inertia in the dynamic case} \label{appAesti}

The rotational dynamical equations discussed in the previous section have a direct parallel with the orbital dynamical equations used in software like GEODYN.  The orbital dynamical differential equations have associated variational differential equations.  The orbital variational equations are derived by differentiating the orbital dynamical equations by each force model parameter (including the initial states).  They yield partial derivatives of the orbital state at any time with respect to the initial state and also with respect to force model parameters, e.g., gravitational coefficients.  Likewise, the rotational dynamical equations have associated variational equations yielding partial derivatives of the orientation angles at any time with respect to the initial angular state (including angular rates) and also with respect to the other force model parameters, e.g., the moments of inertia.  As discussed in Sections~\ref{sec:orientationrepresentation} and \ref{subsec:assmeaspds}, this double set of variational equations needs to be linked in order to include the orientation and moment of inertia parameters as force parameters when considering tracking measurements of the OSIRIX-REx spacecraft.

As with the computations in the previous section, we cast the rotational variational equations in terms of $\mathbf{R}(t)$ and $\boldsymbol{\omega}(t)$, or $\boldsymbol{\zeta}(t)$, and not directly in terms of angles.  Thus, from equation~\ref{eq6ad}, the partial derivative of $\boldsymbol{\zeta}(t)$ with respect to $\boldsymbol{\zeta}_0$ (the initial rotational state) at time $t$, denoted as $\boldsymbol{\Phi}(t)$, satisfies
\begin{equation}
\boldsymbol{\dot{\Phi}}(t)={\partial\mathbf{f}\over\partial\boldsymbol{\zeta}}(t) \boldsymbol{\Phi}(t),\quad\boldsymbol{\Phi}(t_0)=\mathbf{I}, \label{eq9ad}
\end{equation}
where $\mathbf{I}$ is a $12\times 12$ identity matrix.  Likewise, the partial derivative of $\boldsymbol{\zeta}(t)$ with respect to $\mathbf{s}$ (the independent moments of inertia) at time $t$, denoted as $\boldsymbol{\Psi}(t)$, satisfies
\begin{equation}
\boldsymbol{\dot{\Psi}}(t)={\partial\mathbf{f}\over\partial\boldsymbol{\zeta}}(t) \boldsymbol{\Psi}(t)+{\partial\mathbf{f}\over\partial\mathbf{s}}(t),\quad\boldsymbol{\Psi}(t_0)=\mathbf{0}, \label{eq10ad}
\end{equation}
where $\mathbf{0}$ is a $12\times 6$ zero matrix.  Note that equations~\ref{eq9ad} and \ref{eq10ad} differ in two aspects:  the initial conditions and the use of explicit partial derivatives in the latter.  Equations~\ref{eq6ad}, \ref{eq9ad}, and \ref{eq10ad} can be integrated together to yield $\boldsymbol{\zeta}(t)$ and its changes with respect to $\boldsymbol{\zeta}_0$ and $\mathbf{s}$ and can be chained with partial derivatives based on equations~\ref{eq1ad} and \ref{eq4ad} to provide partial derivatives of $\boldsymbol{\zeta}_0$ with respect to the orientation angles and their rates at the initial time $t_0$.

Finally, it should be mentioned that Bennu is located in a low-torque region of space such that $\boldsymbol{\tilde{\tau}}(t)$ is likely to be very small or negligible.  In this case, equation~\ref{eq5ad} becomes
\begin{equation}
\boldsymbol{\dot{\omega}}(t)=-\mathbf{I}_c^{-1}\left[{\boldsymbol{\omega}(t)\times\mathbf{I}_c \boldsymbol{\omega}(t)}\right], \label{eq11ad}
\end{equation}
which means that $\boldsymbol{\dot{\omega}}(t)$ is invariant to a scale factor on $\mathbf{I}_c$.  Thus, any attempt to independently estimate all elements of $\mathbf{s}$ will fail.  In that case, additional constraints must be placed on $\mathbf{s}$, such as fixing the value of one of the elements of $\mathbf{s}$ and solving for the remainder, or fixing $I_{xx}^2+I_{yy}^2+I_{zz}^2$ to a given value, as suggested in Section~\ref{sec:sensitivity}.

\bibliographystyle{spbasic}      


\end{document}